# Implicit Copulas from Bayesian Regularized Regression Smoothers

Nadja Klein and Michael Stanley Smith⋆

University of Melbourne

May 14, 2018

⋆ Nadja Klein is an Alexander von Humboldt Feodor-Lynen Post-doctoral fellow hosted at the University of Melbourne. Michael Stanley Smith is Professor of Management (Econometrics) at Melbourne Business School, University of Melbourne. Correspondence should be directed to Michael Stanley Smith at Melbourne Business School, 200 Leicester Street, Carlton, VIC, 3053, Australia. Email: mike.smith@mbs.edu. Nadja Klein gratefully acknowledges funding from the Alexander von Humboldt foundation. We thank MutualArt for providing the print price data.

# Implicit Copulas from Bayesian Regularized Regression Smoothers

## Abstract

We show how to extract the implicit copula of a response vector from a Bayesian regularized regression smoother with Gaussian disturbances. The copula can be used to compare smoothers that employ different shrinkage priors and function bases. We illustrate with three popular choices of shrinkage priors — a pairwise prior, the horseshoe prior and a g prior augmented with a point mass as employed for Bayesian variable selection — and both univariate and multivariate function bases. The implicit copulas are high-dimensional, have flexible dependence structures that are far from that of a Gaussian copula, and are unavailable in closed form. However, we show how they can be evaluated by first constructing a Gaussian copula conditional on the regularization parameters, and then integrating over these. Combined with non-parametric margins the regularized smoothers can be used to model the distribution of non-Gaussian univariate responses conditional on the covariates. Efficient Markov chain Monte Carlo schemes for evaluating the copula are given for this case. Using both simulated and real data, we show how such copula smoothing models can improve the quality of resulting function estimates and predictive distributions.

# 1 Introduction

A popular way to estimate a smooth unknown function from noisy data is to approximate it with a linear combination of basis functions in a regression with coefficients that are regularized; for example, see Wahba (1990) and Ruppert et al. (2003). We refer to such an approximation as a regularized regression smoother. In a Bayesian context, the regularization term arises from adopting a shrinkage prior for the coefficients. When the response is Gaussian, conditional on the function, it is common to adopt a conjugate conditionally Gaussian shrinkage prior. Examples include (but are not limited to) the pairwise priors of penalized splines (Lang and Brezger, 2004; Fahrmeir and Kneib, 2011), the horseshoe prior (Carvalho and Polson, 2010; Polson and Scott, 2010), Bayesian variable selection priors (George and McCulloch, 1993; Smith and Kohn, 1996) and the Bayesian lasso (Park and Casella, 2008; Hans, 2009). In this paper we show how to extract the implicit copula of the distribution of the response vector from such a smoother. This captures the dependence structure between the elements of the vector. It is constructed by a process called inversion by Nelson (2006, Sec. 3.1), and is a function of the covariates. The implicit copula can be used to compare the smoothing properties of different combinations of priors and function bases. Moreover, combining them with non-parametric marginal distributions results in new regularized regression smoothers for non-Gaussian data. We call these 'copula smoothers', and they have exactly the same dependence structure as that of the original smoother, but are substantially more flexible. As such, the provide an alternative approach to semi-parametric distributional regression (Rigby and Stasinopoulos, 2005; Klein et al., 2015; Wood et al., 2016) for a univariate response.

We first derive the copula with the basis coefficients integrated out. This is a Gaussian copula (Song, 2000) with a correlation matrix that is a function of the covariate values and the hyper-parameters. The latter can include parameters that allow the basis to be of varying dimension. We then integrate over the hyper-parameters to obtain the implicit copula of the regularized regression smoother, which is not a Gaussian copula and unavailable in closed form. In a Bayesian context, the integration can be done with respect to the prior or posterior of the hyper-parameters. In either case, we stress here that the resulting implicit copula has a dependence structure that is very different from that of a Gaussian copula – something we illustrate in our empirical work. The implicit copula density can be expressed as an integral that can be computed readily using Bayesian methods – even when the dimension of the copula is high. The approach of representing the implicit copula as a mixture of Gaussian copulas greatly simplifies its computation, compared to direct evaluation as in Smith and Maneesoonthorn (2016) for state space models.

Three shrinkage priors for the basis coefficients are considered in detail: an autoregressive prior, a horseshoe prior (Carvalho and Polson, 2010) and a g prior augmented with point mass (Smith and Kohn, 1996). These are combined with matching bases, including a B-spline, augmented Fourier and regression spline bases for univariate functions, and additive or radial bases for multivariate functions. We show how to compute dependence metrics (such as Spearman's rho or quantile dependence) between the response variable at two different covariate values. Varying these covariate values produces a surface of dependence metric values that characterize the level of smoothing of the regression smoother. Different combinations of shrinkage prior and basis result in large differences in these surfaces.

When employing the proposed implicit copulas to construct the new copula smoothers, we outline efficient Markov chain Monte Carlo (MCMC) schemes to estimate the posteriors for each choice of shrinkage prior. The regression function is the mean of the response, conditional on the covariate values, and we show how to estimate it using its posterior mean. We also show how to compute the Bayesian predictive density of the response as a function of the covariates. A simulation study illustrates the effectiveness of the copula smoother, and demonstrates that when the data generating process is not conditionally Gaussian, the copula smoother is more accurate – particularly its predictive density.

The approach is extended to construct the implicit copula of an additive regression smoother. However, when this copula is employed with non-parametric margins to model non-Gaussian data, the response is no longer additive in the covariates. Therefore, standard partial residuals (Hastie and Tibshirani, 1990) cannot be computed, and we outline how to compute both function estimates and partial residuals on the domain of the regularized smoother instead. To illustrate, the model is applied to the widely studied Boston housing dataset of Harrison and Rubinfeld (1978). The copula smoother captures the non-Gaussian marginal distribution, improving the accuracy of the predictive density.

The method can also be employed using a radial basis (Powell, 1987) for multivariate functions. We consider two different radial basis functions, a thin-plate spline and a Gaussian kernel, along with the horseshoe and Bayesian variable selection shrinkage priors. These are used to model the logarithm of the price of $n = 11,375$ prints sold at fine art auctions as a function of two covariates. The distribution of the response is non-Gaussian and captured using a non-parametric margin. The bivariate function estimates show the viability of using the copula smoother, even when the $n$-dimensional copula is of very high dimension.

The implicit copulas of elliptical (Fang et al., 2002; McNeil et al., 2005) and skew-elliptical (Demarta and McNeil, 2005; Smith et al., 2012) distributions are employed widely. More recently, interest has grown in computing the implicit copulas of response values (which we call pseudo-response as they are not observed directly) of more complex statistical models. Examples include implicit copulas of Gaussian vector autoregressions (Smith and Vahey, 2016), factor models (Murray et al., 2013; Oh and Patton, 2017) and state space models (Smith and Maneesoonthorn, 2016). In addition, Pitt et al. (2006) and Murray et al. (2013) consider using shrinkage priors for the correlation matrix of a Gaussian copula. However, as far as we are aware, ours is the first paper to consider constructing the implicit copula of the regularized regression smoothing models of the type considered here. Acar et al. (2011) and Craiu and Sabeti (2012) consider copulas with dependence parameters that are functions of one or more covariates. However, these are low-dimensional copulas capturing the dependence between two or more response variables, and are very different from those considered here. In contrast, our implicit copulas capture the dependence between multiple values of a single response variable as a function of the covariates, with the dependence structure inherited from the regularized regression smoother.

In the machine learning literature Gaussian process-based regression smoothers — such as support or relevance vector machines (Tipping, 2001) — are a popular alternative to regularized smoothers of the type considered here. While a number of authors extend Gaussian processes by constructing their implicit copulas (Wauthier and Jordan, 2010; Wilson and Ghahramani, 2010), we are unaware of any work constructing the implicit copula of vector machines. Moreover, these copulas are Gaussian copulas, whereas the implicit copulas constructed here are not. Gaussian processes have also been used as building blocks along with conditional copulas to model non-Gaussian regression or time series data (Hernández-Lobato et al., 2013; Levi and Craiu, 2016). However, these approaches employ low-dimensional closed form parametric copulas. In contrast, the implicit copulas proposed here are high-dimensional and unavailable in closed form, and are very different.

The rest of this paper is structured as follows. Section 2 outlines our approach to constructing the implicit copula; both in general and for the three considered in detail. Section 3 outlines how to employ the proposed copula with arbitrary margins to model non-Gaussian data. Section 4 contains the simulation study. Section 5 extends our copula to both additive models and radial bases, and illustrates using the two real datasets. Section 6 concludes.

## 2 Implicit Copula

In this section, we explain our approach for constructing the copula of a regularized regression model for the pseudo response with a single covariate. It is extended to the case of multiple covariates in Section 5. To do so we first construct the Gaussian copula conditional on the regularization parameters, and then construct the implicit copula as an integral over these.

### 2.1 The General Idea

Consider the regression model

$$\tilde{Z}_i = \tilde{m}(x_i) + \varepsilon_i, \quad \text{for } i = 1, \ldots, n \tag{1}$$

for a pseudo response $\tilde{Z}_i$, where $\tilde{m}$ is an unknown univariate function, $x_i$ is a covariate value, and $\varepsilon_i$ is distributed independently $N(0, \sigma^2)$. It is popular to model $\tilde{m}$ as a linear combination of $p$ basis functions $b_1, \ldots, b_p$, so that $\tilde{m}(x) = \sum_{j=1}^{p} \beta_j b_j(x)$. In this case, (1) can be rewritten as the linear model

$$\tilde{\bm{Z}} = B\bm{\beta} + \bm{\varepsilon}, \tag{2}$$

for $\tilde{\bm{Z}} = (\tilde{Z}_1, \ldots, \tilde{Z}_n)' \in \mathbb{R}^n$, with $\bm{\varepsilon} \sim \mathrm{N}(0, \sigma^2 I)$. The $(n \times p)$ design matrix $B$ has $i$th row $\bm{b}_i' = (b_1(x_i), \ldots, b_p(x_i))$ evaluated at $x_i$. There are many bases used in practise, and we consider three common choices here: regression splines (Friedman, 1991), B-splines (De Boor, 1978) and an augmented Fourier basis.

In the Bayesian literature, priors are employed on $\bm{\beta} = (\beta_1, \ldots, \beta_p)'$ to allow for a data-driven level of shrinkage to provide a smooth, but flexible, estimate of $\tilde{m}$. We follow this approach and employ the prior

$$\bm{\beta}|\bm{x}, \sigma^2, \bm{\theta}, \bm{\gamma} \sim N(\bm{0}, \sigma^2 P(\bm{\theta})^{-1}), \qquad (3)$$

where the precision matrix $P(\bm{\theta})$ is of full rank. The parameters $\bm{\theta}$ are shrinkage parameters, while $\bm{\gamma}$ are further parameters that allow for the basis to be of varying dimension (which we discuss later). The matrix $P$ may also be a function of the covariate vector $\bm{x} = (x_1, \ldots, x_n)'$. In Section 2.2 we consider three different priors of this form.

Conditional on $(\bm{x}, \sigma^2, \bm{\theta}, \bm{\gamma})$, the prior of $\bm{\beta}$ is conjugate, so that it can be integrated out to give

$$\tilde{\bm{Z}}|\bm{x}, \sigma^2, \bm{\theta}, \bm{\gamma} \sim \mathrm{N}(\bm{0}, \sigma^2 (I - B\Omega B')^{-1}), \qquad (4)$$

with $\Omega = (B'B + P(\bm{\theta}))^{-1}$. Application of the Woodbury formula gives

$$(I - B\Omega B')^{-1} = I + B\left[B'B + P(\bm{\theta}) - B'B\right]^{-1} B' = I + BP(\bm{\theta})^{-1}B',$$

with $i$th diagonal element $1 + \bm{b}_i' P(\bm{\theta})^{-1} \bm{b}_i$. Therefore, the $i$th margin of this distribution is $\tilde{Z}_i|\bm{x}, \sigma^2, \bm{\theta}, \bm{\gamma} \sim \mathrm{N}(0, \sigma^2(1 + \bm{b}_i' P(\bm{\theta})^{-1} \bm{b}_i))$.

The copula of the distribution at (4) is called a Gaussian copula (Song, 2000), and is constructed by standardizing the distribution to have zero mean and unit variances. To do so here, we set $\bm{Z} = \sigma^{-1} S(\bm{x}, \bm{\theta}, \bm{\gamma}) \tilde{\bm{Z}}$, where $S(\bm{x}, \bm{\theta}, \bm{\gamma}) = \mathrm{diag}(s_1, \ldots, s_n)$ is a diagonal scaling matrix with elements $s_i = [1 + \bm{b}_i' P(\bm{\theta})^{-1} \bm{b}_i]^{-1/2}$. With this standardization, the regression at (1) can be rewritten as

$$Z_i = m(x_i) + \frac{s_i}{\sigma} \varepsilon_i, \qquad (5)$$

where $m(x_i) = (s_i/\sigma) \bm{b}_i' \bm{\beta}$ and both $s_i$ and $\bm{b}_i$ are functions of $x_i$. The conditional distribution of the standardized vector $\bm{Z} = (Z_1, \ldots, Z_n)'$ is then

$$\bm{Z}|\bm{x}, \bm{\beta}, \sigma^2, \bm{\theta}, \bm{\gamma} \sim \mathrm{N}\left(\frac{S(\bm{x}, \bm{\theta}, \bm{\gamma})}{\sigma} B\bm{\beta}, S(\bm{x}, \bm{\theta}, \bm{\gamma}) S(\bm{x}, \bm{\theta}, \bm{\gamma})'\right). \qquad (6)$$

Integrating out $\boldsymbol{\beta}$ as before, gives the unconditional (on $\boldsymbol{\beta}$) distribution of $\boldsymbol{Z}$, which we summarize in the following Theorem.

**Theorem 1.**

Let $\tilde{\boldsymbol{Z}}$ follow the linear model at (2), with the prior for $\boldsymbol{\beta}$ as given at (3). Then:

(i) The joint distribution $\boldsymbol{Z}|\boldsymbol{x}, \sigma^2, \boldsymbol{\theta}, \boldsymbol{\gamma} \sim N(\boldsymbol{0}, R(\boldsymbol{x}, \boldsymbol{\theta}, \boldsymbol{\gamma}))$ with

$$R(\boldsymbol{x}, \boldsymbol{\theta}, \boldsymbol{\gamma}) = S(\boldsymbol{x}, \boldsymbol{\theta}, \boldsymbol{\gamma})(I - B\Omega B')^{-1}S(\boldsymbol{x}, \boldsymbol{\theta}, \boldsymbol{\gamma})'. \tag{7}$$

(ii) The marginal distributions $Z_i|\boldsymbol{x}, \sigma^2, \boldsymbol{\theta}, \boldsymbol{\gamma} \sim N(0, 1)$ for $i = 1, \ldots, n$.

(iii) The copula of both $\tilde{\boldsymbol{Z}}$ and $\boldsymbol{Z}$, conditional on $(\boldsymbol{x}, \boldsymbol{\theta}, \boldsymbol{\gamma})$, is a Gaussian copula with copula function $C(\boldsymbol{u}|\boldsymbol{x}, \boldsymbol{\theta}, \boldsymbol{\gamma}) = \Phi_n\left(\Phi_1^{-1}(u_1), \ldots, \Phi_1^{-1}(u_n); \boldsymbol{0}, R(\boldsymbol{x}, \boldsymbol{\theta}, \boldsymbol{\gamma})\right)$, where $\boldsymbol{u} = (u_1, \ldots, u_n)'$, while $\Phi_n(\cdot; \boldsymbol{0}, R)$ and $\Phi_1$ are the distribution functions of $N_n(\boldsymbol{0}, R)$ and $N(0, 1)$ distributions, respectively.

(iv) The corresponding copula density is

$$c(\boldsymbol{u}|\boldsymbol{x}, \boldsymbol{\theta}, \boldsymbol{\gamma}) = \frac{p(\boldsymbol{z}|\boldsymbol{x}, \sigma^2, \boldsymbol{\theta}, \boldsymbol{\gamma})}{\prod_{i=1}^n p(z_i|\boldsymbol{x}, \sigma^2, \boldsymbol{\theta}, \boldsymbol{\gamma})} = \frac{\phi_n(\boldsymbol{z}; \boldsymbol{0}, R(\boldsymbol{x}, \boldsymbol{\theta}, \boldsymbol{\gamma}))}{\prod_{i=1}^n \phi_1(z_i)}, \tag{8}$$

where $z_i = \Phi^{-1}(u_i)$, $\boldsymbol{z} = (z_1, \ldots, z_n)'$ and $\phi_n(\cdot; \boldsymbol{0}, R)$ and $\phi_1$ are the densities of $N_n(\boldsymbol{0}, R)$ and $N(0, 1)$ distributions, respectively.

We make five observations concerning Theorem 1 above. First, $\sigma^2$ does not feature in the expression for the copula function or density and is therefore unidentified, so that we simply set it to 1 throughout the rest of the paper. This is because the copula is invariant to the scale of $Z_i$. Second, if a non-conjugate prior is used for $\boldsymbol{\beta}|\boldsymbol{x}, \boldsymbol{\theta}, \boldsymbol{\gamma}$, then the implicit copula above would not be a Gaussian copula. Third, if an improper prior is employed for $\boldsymbol{\beta}$ — such as those popular in the Bayesian spline literature (Speckman and Sun, 2003; Lang and Brezger, 2004) — then the distribution $\boldsymbol{Z}|\boldsymbol{x}, \boldsymbol{\theta}, \boldsymbol{\gamma}$ is also improper, and the copula is undefined. Therefore, we only employ strictly proper priors here. Fourth, while the copula is $n$-dimensional (so that it can be of very high dimension), the matrix $R$ at (7) is parsimonious because it is a function of $(\boldsymbol{\theta}, \boldsymbol{\gamma})$. In the next subsection we give expressions for $R$ for the three shrinkage priors considered in detail. Last, while the copula at (8) is Gaussian, mixing over the distribution $\pi(\boldsymbol{\theta}, \boldsymbol{\gamma})$ results in a non-Gaussian copula that cannot in general be expressed in closed form, as summarized in the following corollary.

**Corollary 1.** *If $\tilde{\bm{Z}}$ follows the linear model at (2), with the prior for $\bm{\beta}$ given at (3), and $\pi(\bm{\theta}, \bm{\gamma})$ is a proper density, then*

$$c_\pi(\bm{u}|\bm{x}) = \int \int c(\bm{u}|\bm{x}, \bm{\theta}, \bm{\gamma}) \pi(\bm{\theta}, \bm{\gamma}) \mathrm{d}(\bm{\theta}, \bm{\gamma})$$

*is also a copula density, and is not a Gaussian copula.*

The proof of Corollary 1 can be found in Appendix A. The corresponding copula function is denoted as $C_\pi(\bm{u}|\bm{x}) = \int \int C(\bm{u}|\bm{x}, \bm{\theta}, \bm{\gamma}) \pi(\bm{\theta}, \bm{\gamma}) \mathrm{d}(\bm{\theta}, \bm{\gamma})$. In this paper, we consider both the prior $\pi_0(\bm{\theta}, \bm{\gamma})$ and the posterior $p(\bm{\theta}, \bm{\gamma}|\bm{y})$ densities for $\pi(\bm{\theta}, \bm{\gamma})$. When a regularized smoother is fit to data, it is this mixture copula that captures the dependence structure of the resulting data distribution. Evaluation of (and generation from) $c_\pi$ and $C_\pi$ can be undertaken efficiently by Monte Carlo simulation, as we show later.

Representation of $C_\pi$ as a mixture of Gaussian copulas greatly simplifies its computation. In contrast, $C_\pi$ is much harder to compute via inversion of the distribution $\tilde{\bm{Z}}|\bm{x}$ directly. This is because the marginal distribution function of $\tilde{Z}_i|\bm{x}$ is

$$F(\tilde{z}_i|\bm{x}) = \int \Phi_1\left(\tilde{z}_i; 0, (1 + \bm{b}_i' P(\bm{\theta})^{-1} \bm{b}_i)\right) \pi(\bm{\theta}, \bm{\gamma}) \mathrm{d}(\bm{\theta}, \bm{\gamma}),$$

where the integral typically requires computation via numerical methods. The direct inversion approach requires evaluation of the corresponding quantile functions $\tilde{z}_i = F^{-1}(u_i|\bm{x})$, for $i = 1, \ldots, n$, which is prohibitively slow for large sample sizes. Instead, the 'conditioning trick' suggested here makes computation of the copula much faster, as shown in Section 5 for two high-dimensional cases.

## 2.2 Three Implicit Copulas

We construct implicit copulas using three popular shrinkage priors for $\bm{\beta}$. Each prior is of the form at (3), and is usually matched with specific bases. We discuss each in further detail below and summarize them in Tab. 1.

### 2.2.1 P-Spline Copula (PSC)

There is an extensive literature on Bayesian P-splines that employ differenced priors, also called random walk priors (Fahrmeir and Lang, 2001; Lang and Brezger, 2004). However, these are improper, so that $Z|x,\theta,\gamma$ with $\beta$ integrated out is also, and the copula at (8) undefined. Therefore, we instead employ a first order stationary autoregression $\beta_i|\beta_{i-1} \sim N(\psi\beta_{i-1}, \tau^2)$, which approximates a first order random walk when $\psi \to 1$. For this prior, $\gamma = \emptyset$, $\theta = \{\psi, \tau\}$, and $P(\theta) = (\tau^2)^{-1}\Delta(\psi)'\Delta(\psi) = (\tau^2)^{-1}P(\psi)$ is a full rank band one matrix, with upper triangular Cholesky factor $\Delta(\psi)$. Following Eilers and Marx (1996), we match this prior with a B-spline basis of degree $l = 3$ (i.e. a cubic B-spline) with $m + 2l$ equally-spaced knots, where $m$ is the number of inner knots. In our empirical work, we set $m$ to values between 20 and 30, which is a typical choice in applied analysis, resulting in a dimension of $m + l - 1$ for $\beta$.

For the prior $\pi_0(\theta)$ we assume $\psi$ and $\tau^2$ are independent, with $\psi \sim$ Uniform$(0.01, 0.99)$, so that there is positive dependence between coefficients and $P(\psi)$ is full rank. For $\tau^2$, several proper priors have been studied in the literature (Gelman, 2005). Klein and Kneib (2016) study the issue in depth, and recommend scale-dependent priors motivated from the general concept of penalized complexity priors. Following these authors, we employ the Weibull distribution with scale parameter $b_{\tau^2} = 2.5$ in our empirical work. From Theorem 1, the correlation matrix

$$R(x,\theta) = S(x,\theta)(I + \tau^2 BP(\psi)^{-1}B')S(x,\theta),$$

and we label the implicit copula 'PSC'.

Last, note that the $\psi$ and $\tau^2$ control different aspects of the dependence structure, as illustrated in Section 2.3. Moreover, higher order autoregressive priors for $\beta$ can also be used, akin to the popular higher order random walks (Fahrmeir and Kneib, 2011).

### 2.2.2 Horseshoe Copula (HSC)

The horseshoe prior (Carvalho and Polson, 2010) is attractive due to its robustness, adaptivity to sparseness patterns and analytical properties (Polson and Scott, 2010; Bhatacharya et al., 2016). It is a scale mixture, where $\beta_j|\lambda_j \sim N(0, \lambda_j^2)$, with local shrinkage parameters $\lambda_j$, $\pi_0(\lambda_j|\tau) =$ Half-Cauchy$(0, \tau)$ and common scale $\tau$, $\pi_0(\tau) =$ Half-Cauchy$(0, 1)$. With this prior $\gamma = \emptyset$, $\theta = \{\lambda, \tau\}$, with $\lambda = (\lambda_1, \ldots, \lambda_p)'$, while the correlation matrix

$$R(x,\theta) = S(x,\theta)(I + B \operatorname{diag}(\lambda_1, \ldots, \lambda_p)^2 B')S(x,\theta).$$

While we are unaware of any previous usage of the horseshoe prior for regularized smoothing, the localized shrinkage of the prior makes it an attractive choice. Here, we employ the prior with two univariate bases. The first is the same B-spline basis employed for the PSC, while a second is the augmented Fourier basis of $2K$ basis terms $\{\sin(k\pi x), \cos(k\pi x)\,;\, k = 1, \ldots, K\}$, where the covariate is scaled to $[0, 1]$ and we typically set $K = 10$ in our empirical work. We label this copula 'HSC'.

### 2.2.3 Bayesian Variable Selection Copula (BVSC)

For this prior, $\boldsymbol{\theta} = \emptyset$, so that we drop reference to it when discussing this implicit copula. Spike-and-slab priors are popular in the Bayesian variable selection literature; see Clyde and George (2004) for a review. They allow for bases of varying dimension, with $\boldsymbol{\gamma} = (\gamma_1, \ldots, \gamma_p)'$ a vector of binary indicators ($\gamma_i \in \{0, 1\}$) denoting whether, or not, each basis term is included or omitted from $p$ candidates. Let $p_\gamma = \sum_{i=1}^p \gamma_i$, and at (3) denote $\boldsymbol{\beta}, B$ and $P$ as $\boldsymbol{\beta}_\gamma, B_\gamma$ and $P_\gamma$, respectively. We adopt the g prior for the included terms, where $\boldsymbol{\beta}_\gamma | \boldsymbol{\gamma} \sim N(\mathbf{0}, P_\gamma^{-1})$, with $P_\gamma^{-1} = c(B_\gamma' B_\gamma)^{-1}$ and $c = 100$ as in Smith and Kohn (1996). Substituting $P_\gamma$ into (7), the correlation matrix

$$R(\boldsymbol{x}, \boldsymbol{\gamma}) = S(\boldsymbol{x}, \boldsymbol{\gamma})(I + cB_\gamma(B_\gamma' B_\gamma)^{-1} B_\gamma') S(\boldsymbol{x}, \boldsymbol{\gamma}),$$

$\Omega = \frac{c}{1+c}(B_\gamma' B_\gamma)^{-1}$, and $\boldsymbol{b}_{\gamma,i}$ is the $i$th row of $B_\gamma$. Note that for this prior $s_i = (1 + c\boldsymbol{b}_{\gamma,i}'(B_\gamma' B_\gamma)^{-1} \boldsymbol{b}_{\gamma,i})^{-1/2}$, and is a function of all elements of $\boldsymbol{x}$, not just $x_i$.

We use the prior mass function $\pi_0(\boldsymbol{\gamma}) = \text{Beta}(p - p_\gamma + 1, p_\gamma + 1)$. This has been used extensively in the Bayesian selection literature and accounts for the multiplicity of the $2^p$ possible configurations of $\boldsymbol{\gamma}$ (Scott and Berger, 2010). It implies a uniform distribution on $\pi_0(p_\gamma) = 1/(p+1)$ and Bernoulli margins $\Pr(\gamma_i = 1) = 1/2$. We employ this prior with the cubic regression spline basis $\{x, x^2, x^3, (x - k_1)_+^3, \ldots, (x - k_K)_+^3\}$, where $\{a\}_+^3 = \min(0, a^3)$ and $k_1, \ldots, k_K$ are knots chosen to follow the empirical distribution of the covariate with $K = 25$. We label this implicit copula 'BVSC'.

## 2.3 Dependence Structure

Consider two new covariate values $\boldsymbol{x}_0 = (x_{0,1}, x_{0,2})'$ with corresponding pseudo response values $Z_{0,1}, Z_{0,2}$ for the standardized response at (6). Denote the vector of these two values combined with the other $n$ covariate observations as $\bm{x}^+ = (\bm{x}_0', \bm{x}')'$. We use metrics of pairwise dependence between $Z_{0,1}$ and $Z_{0,2}$ to measure the dependence structure of the implicit copula. Possible metrics include quantile dependence and Kendall's tau (Nelson, 2006, Chapter 5), but we illustrate here using Spearman's correlation.

From Theorem 1, $C(\bm{u}|\bm{x},\bm{\theta},\bm{\gamma})$ is a Gaussian copula, and the Spearman correlation between $Z_{0,1}$ and $Z_{0,2}$ for this copula is

$$\rho^s(x_{0,1}, x_{0,2}|\bm{x},\bm{\theta},\bm{\gamma}) = \frac{6}{\pi}\arcsin(r_{12}(\bm{x}^+,\bm{\theta},\bm{\gamma})),$$

where $r_{12}(\bm{x}^+,\bm{\theta},\bm{\gamma})$ is the off-diagonal element giving the pairwise correlation between $Z_{0,1}$ and $Z_{0,2}$ in the $(n+2)\times(n+2)$ matrix $R(\bm{x}^+,\bm{\theta},\bm{\gamma})$. For the PSC and HSC implicit copulas, it is straightforward to show that $r_{12}$ is a function of only $(x_{0,1}, x_{0,2})$ and not $\bm{x}$, so that $\rho^s$ is also. However, for the BVSC implicit copula $r_{12}$ depends on all elements of $\bm{x}^+$ because each element of the diagonal scaling matrix $S$ does so also. It is this feature that makes the smoothing locally adaptive for this prior, as discussed further in Section 2.4.

The same dependence metrics for the mixture copula $C_\pi$ at Corollary 1 can be computed via simulation. For example, the Spearman's correlation between $Z_{0,1}$ and $Z_{0,2}$ from this copula is

$$\rho^s_\pi(x_{0,1}, x_{0,2}|\bm{x}) = \int \rho^s(x_{0,1}, x_{0,2}|\bm{x},\bm{\theta},\bm{\gamma})\pi(\bm{\theta},\bm{\gamma})\mathrm{d}(\bm{\theta},\bm{\gamma}) \approx \frac{1}{J}\sum_{j=1}^{J} \rho^s(x_{0,1}, x_{0,2}|\bm{x},\bm{\theta}^{[j]},\bm{\gamma}^{[j]}),$$

where $(\bm{\theta}^{[j]},\bm{\gamma}^{[j]})' \sim \pi(\bm{\theta},\bm{\gamma})$ and $J$ is the total number of iterates. Simulating from $\pi$ is typically straightforward when it is the prior distribution, and can be achieved using the MCMC methods in Section 3 when it is the posterior.

## 2.4 Illustration

To illustrate the dependence structure of our proposed copulas, we first consider the PSC with $\bm{\theta} = \{\psi, \tau^2\}$. Fig. 2 shows $\rho^s$ as a function of $(x_{0,1} - x_{0,2})$, where in panel (a) $\psi = 0.5$ and $\tau^2 \in \{0.01, 0.1, 0.5, 1, 10, 100\}$, and in panel (b) $\tau^2 = 1$ and $\psi \in \{0.1, 0.25, 0.5, 0.75, 0.9, 0.95\}$. This reveals that $\tau^2$ determines the overall level of dependence between $Z_{0,1}$ and $Z_{0,2}$, while $\psi$ determines how quickly $\rho^s$ decreases as $|x_{0,1} - x_{0,2}|$ increases. The dependence is symmetric around $(x_{0,1} - x_{0,2}) = 0$.

We next compare the dependence structure of the three (non-Gaussian) implicit copulas $C_\pi$ for the prior $\pi = \pi_0$. Because $\rho_\pi^s$ is a function of $\boldsymbol{x}$ for the BVSC, $n = 200$ covariate values are generated from a $\chi^2$ distribution and scaled to $[0, 1]$. Fig. 1(a) shows a histogram of these values. We then compute $\rho_\pi^s$ over a bivariate grid for $(x_{0,1}, x_{0,2})$ on the unit square, with $J = 10,000$ iterates simulated from the priors $\pi_0$ for each case. Fig. 3 plots $\rho_\pi^s$ as surfaces on the left-hand side for four cases: (a) PSC with a B-spline basis, (c) HSC with a B-spline basis, (e) HSC with an augmented Fourier basis, and (g) BVSC with a regression spline basis.

We make five observations. First, in each case $\rho_\pi^s$ is highest as $|x_{0,1} - x_{0,2}| \to 0$. This is expected for any effective smoother, because response values should be more dependent when their covariate values are closer. Second, even though the function bases are identical in panels (a) and (c), the level of smoothing is higher with the PSC than HSC. Clearly, the shrinkage prior employed for $\boldsymbol{\beta}$ has a strong impact on the dependence structure. Third, even though the prior for $\boldsymbol{\beta}$ is the same in panels (c) and (e), the bases employed are different, which also has a large effect on the dependence structure. Fourth, $\rho_\pi^s$ is non-monotonic in $|x_{0,1} - x_{0,2}|$ for the augmented Fourier basis, with 'ripples' observed. This is because the basis terms are non-monotonic in the covariate value. Fifth, the BVSC is the only case where the $n$ values of $\boldsymbol{x}$ have an impact on $\rho_\pi^s$, which can be seen in panel (g). There is higher smoothing for values of $x_{0,1}$ and $x_{0,2}$ close to 1 (i.e. in the top right-hand corner), and lower smoothing for values around 0.3. This is local adaptivity, with higher levels of smoothing occurring where the design is sparse, and lower levels of smoothing where the design is dense.

## 3 Copula Smoother

The main application of our proposed copula is in conjunction with arbitrary marginal distributions to model non-Gaussian regression data. In this section we outline this model, and Bayesian methods to estimate the copula parameters, regression function and predictive distributions.

## 3.1 Observational Model and Likelihood

Let $\bm{Y} = (Y_1, \ldots, Y_n)'$ be $n$ observations on a continuous response, with covariate values $\bm{x}$. We assume throughout that $Y_i|x_i$ has a distribution function $F_Y$ and density $p_Y$ that does not vary with $i$. The joint density of $\bm{Y}|\bm{x}$ is

$$p(\bm{y}|\bm{x}) = c_\pi \left(F_Y(y_1), \ldots, F_Y(y_n)|\bm{x}\right) \prod_{i=1}^n p_Y(y_i),$$

where $c_\pi$ is the copula density at Corollary 1 with $\pi(\bm{\theta}, \bm{\gamma}) = p(\bm{\theta}, \bm{\gamma}|\bm{y})$ the posterior distribution. We call this model a 'copula smoother', because all regression smoothing is introduced through the copula only, and not the margin $p_Y$.

From Theorem 1, the likelihood conditional on $\bm{\theta}, \bm{\gamma}$, but with $\bm{\beta}$ marginalized out, is

$$p(\bm{y}|\bm{x}, \bm{\theta}, \bm{\gamma}) = p(\bm{z}|\bm{x}, \bm{\theta}, \bm{\gamma}) \prod_{i=1}^n \frac{p_Y(y_i)}{p(z_i|\bm{x}, \bm{\theta}, \bm{\gamma})} = \phi_n(\bm{z}; \bm{0}, R(\bm{x}, \bm{\theta}, \bm{\gamma})) \prod_{i=1}^n \frac{p_Y(y_i)}{\phi_1(z_i)}. \quad (9)$$

For large $n$, direct computation of the $n \times n$ correlation matrix $R$ is computationally infeasible. However, the likelihood conditional on $\bm{\beta}$ is

$$p(\bm{y}|\bm{x}, \bm{\beta}, \bm{\theta}, \bm{\gamma}) = p(\bm{z}|\bm{x}, \bm{\beta}, \bm{\theta}, \bm{\gamma}) \prod_{i=1}^n \frac{p_Y(y_i)}{\phi_1(z_i)} = \phi_n(\bm{z}; SB\bm{\beta}, SS') \prod_{i=1}^n \frac{p_Y(y_i)}{\phi_1(z_i)},$$

which can be evaluated in $O(n)$ operations because $S$ is diagonal. We exploit this observation to propose MCMC schemes below that avoid direct computation of $R$.

## 3.2 Posterior Evaluation

We estimate the marginal density non-parametrically using the adaptive kernel density estimator of Shimazaki and Shinomoto (2010), and use this to compute $z_i = \Phi_1^{-1}(\hat{F}_Y(y_i))$, for $i = 1, \ldots, n$. We use MCMC to compute the posterior of $\bm{\theta}$ augmented with the coefficients $p(\bm{\beta}, \bm{\theta}|\bm{x}, \bm{y})$ for the PSC and HSC, and the posterior $p(\bm{\gamma}|\bm{x}, \bm{y})$ for the BVSC. For the PSC and HSC we generate from the conditional posterior $p(\bm{\beta}|\bm{x}, \bm{\theta}, \bm{y}) = p(\bm{\beta}|\bm{x}, \bm{\theta}, \bm{z})$, which is Gaussian with mean $\bm{\mu}_\beta = \Sigma_\beta B' S^{-1} \bm{z}$ and covariance matrix $\Sigma_\beta = (B'B + P(\bm{\theta}))^{-1}$. The steps required to generate from the conditional posteriors of $\bm{\theta}$ and $\bm{\gamma}$ are outlined separately below for each of the three implicit copulas, while implementation details can be found in Appendix B.

### 3.2.1 PSC

The conditional posterior

$$p(\tau^2|\boldsymbol{x},\boldsymbol{\beta},\psi,\boldsymbol{y}) \propto p(\boldsymbol{z}|\boldsymbol{x},\boldsymbol{\beta},\tau^2,\psi)p(\boldsymbol{\beta}|\tau^2,\psi)\pi_0(\tau^2|b_{\tau^2})$$

$$\propto \prod_{i=1}^{n} \frac{1}{s_i} \exp\left(-\frac{1}{2}(\boldsymbol{z}-SB\boldsymbol{\beta})'(SS')^{-1}(\boldsymbol{z}-SB\boldsymbol{\beta})\right) p(\boldsymbol{\beta}|\tau^2,\psi)\pi_0(\tau^2|b_{\tau^2}),$$

which is not a recognizable distribution. A Metropolis-Hastings step is used to generate $v = \log(\tau^2)$, where a normal distribution matching the mode and curvature is used to approximate its conditional. Note that

$$l_v \equiv \log(p(v|\boldsymbol{x},\boldsymbol{\beta},\psi,\boldsymbol{y})) \propto -\frac{v}{2}\left(\dim(P(\psi))-1\right) - \frac{1}{2\exp(v)}\boldsymbol{\beta}'P(\psi)\boldsymbol{\beta} - \left(\frac{\exp(v)}{b_{\tau^2}}\right)^{\frac{1}{2}}$$

$$-\frac{1}{2}\sum_{i=1}^{n}\log(s_i^2) - \frac{1}{2}\left(\boldsymbol{z}'(SS')^{-1}\boldsymbol{z} - 2\boldsymbol{\beta}'B'S^{-1}\boldsymbol{z}\right).$$

Approximating $l_v$ by a second order Taylor expansion around the current state $v^{(c)}$, and taking the exponent, yields the proposal density $N(\mu_v, \sigma_v^2)$ with $\mu_v = \sigma_v^2 \frac{\partial l_v}{\partial v} + v$ and $\sigma_v^2 = -1/\frac{\partial^2 l_v}{\partial v^2}$. Analytical expressions for the derivatives are given in Appendix B.1.

We transform $\psi$ onto the real line as $\xi = g(\psi) = \log\left((\psi-\epsilon)/(1-\epsilon-\psi)\right)$, with $\epsilon = 0.01$. The log-posterior is

$$l_\xi \equiv \log(p(\xi|\boldsymbol{x},\boldsymbol{\beta},\tau^2,\boldsymbol{y})) \propto \log\left(\frac{\partial \psi}{\partial \xi}\right) + \log(\pi_0(g^{-1}(\xi))) + \log(p(\boldsymbol{z}|\boldsymbol{x},\boldsymbol{\beta},\tau^2,\psi)) + \log(p(\boldsymbol{\beta}|\tau^2,\psi))$$

$$\propto \xi - 2\log(1+\exp(\xi)) + \log(\det(\Delta(g^{-1}(\xi))))$$

$$-\frac{1}{2}\sum_{i=1}^{n}\log(s_i^2) - \frac{1}{2}\left(\boldsymbol{z}'(SS')^{-1}\boldsymbol{z} - 2\boldsymbol{\beta}'B'S^{-1}\boldsymbol{z}\right) - \frac{\boldsymbol{\beta}'P(g^{-1}(\xi))\boldsymbol{\beta}}{2\tau^2}.$$

We generate $\xi$ using a Metropolis-Hastings step in the same fashion as for $v$, but using the derivatives of $l_\xi$ which are given in Appendix B.1. Because both proposals are based on analytical derivatives, they are fast to compute. In our empirical work, the acceptance rates of $v$ and $\xi$ were between 60% and 90%. Last, we found joint updates of $(\tau^2, \psi)$ had prohibitively low acceptance rates.

### 3.2.2 HSC

Both $\tau$ and each element $\lambda_j$ of $\boldsymbol{\lambda}$ are generated separately. Metropolis-Hastings steps with normal approximations as proposals are used as in the PSC case, where

$$\log(p(\log(\lambda_j^2)|\boldsymbol{x},\boldsymbol{\beta},\boldsymbol{\lambda}_{\setminus j},\tau,\boldsymbol{z})) \propto -\frac{1}{2}\sum_{i=1}^{n}\log(s_i^2) - \frac{1}{2}\boldsymbol{z}'(SS')^{-1}\boldsymbol{z} + \boldsymbol{\beta}'B'S^{-1}\boldsymbol{z}$$
$$-\frac{1}{2}\left[\log(\lambda_j^2) + \frac{\beta_j^2}{\lambda_j^2} + 2\log\left(1+\frac{\lambda_j^2}{\tau^2}\right)\right]$$

$$\log(p(\log(\tau)|\boldsymbol{x},\boldsymbol{\lambda},\boldsymbol{z})) \propto -(p-1)\log(\tau) - \log(1+\tau^2) - \sum_{j=1}^{p}\frac{\lambda_j^2}{\tau^2},$$

and $\boldsymbol{\lambda}_{\setminus j}$ denotes $\boldsymbol{\lambda}$ without element $\lambda_j$. The derivatives of the conditional posteriors of $\log(\lambda_j^2)$ and $\log(\tau)$ are given in Appendix B.2. Similar to sampler for the PSC, in our simulations acceptance rates of these steps were around 70% for $\log(\lambda_j^2)$ and above 90% for $\log(\tau)$.

### 3.2.3 BVSC

From (9), the posterior

$$\begin{aligned}p(\boldsymbol{\gamma}|\boldsymbol{x},\boldsymbol{y}) &\propto p(\boldsymbol{y}|\boldsymbol{x},\boldsymbol{\gamma})\pi_0(\boldsymbol{\gamma}) \propto \phi_n(\boldsymbol{z};\boldsymbol{0},R(\boldsymbol{x},\boldsymbol{\gamma}))\pi_0(\boldsymbol{\gamma}) \\ &\propto |R(\boldsymbol{x},\boldsymbol{\gamma})|^{-1/2}\exp\left\{-\frac{1}{2}\left(\boldsymbol{z}'R(\boldsymbol{x},\boldsymbol{\gamma})^{-1}\boldsymbol{z}\right)\right\}\text{Beta}(p-p_\gamma+1,p_\gamma+1) \equiv A(\gamma_i,\gamma_j).\end{aligned}$$

We generate from this posterior using a Gibbs sampler, where $\boldsymbol{\gamma}$ is partitioned into pairs of elements selected at random, and each pair $(\gamma_i,\gamma_j)$ is generated conditional on the other elements $\boldsymbol{\gamma}\setminus(\gamma_i,\gamma_j)$. This involves computing $A(\gamma_i,\gamma_j)$ for the four possible configurations $(\gamma_i,\gamma_j) \in \mathcal{S} \equiv \{(0,0),(0,1),(1,0),(1,1)\}$ for that pair of indicator values. This can be undertaken efficiently as outlined in Appendix B.3, where direct computation of $R$ is avoided. Then we generate from $p((\gamma_i,\gamma_j)|\boldsymbol{\gamma}\setminus(\gamma_i,\gamma_j),\boldsymbol{x},\boldsymbol{y}) = \frac{A(\gamma_i,\gamma_j)}{\sum_{(\tilde{\gamma}_i,\tilde{\gamma}_j)\in\mathcal{S}} A(\tilde{\gamma}_i,\tilde{\gamma}_j)}$. Unlike for the other two implicit copulas, $\boldsymbol{\beta}$ is not generated as part of the MCMC scheme. If it was, the Markov chain would be reducible, which is a well-known issue with models of varying dimension.

## 3.3 Function Estimation

For a new observation $(Y_0,x_0)$ on the response and covariate, to estimate the regression function $f(x_0) \equiv \mathbb{E}(Y_0|x_0)$ we employ the posterior predictive mean

$$\mathbb{E}(Y_0|x_0,\boldsymbol{x},\boldsymbol{y}) = \int \mathbb{E}(Y_0|x_0,\boldsymbol{x},\boldsymbol{\beta},\boldsymbol{\theta},\boldsymbol{\gamma})p(\boldsymbol{\beta},\boldsymbol{\theta},\boldsymbol{\gamma}|\boldsymbol{x},\boldsymbol{y})\mathrm{d}(\boldsymbol{\beta},\boldsymbol{\theta},\boldsymbol{\gamma}).$$

Note that $f$ is different from $m$ in Eq. (5), which is the mean function for the pseudo-response. Let $Z_0 = \Phi_1^{-1}(F_Y(Y_0))$, then the expectation

$$\mathbb{E}(Y_0|x_0, \boldsymbol{x}, \boldsymbol{\beta}, \boldsymbol{\theta}, \boldsymbol{\gamma}) = \mathbb{E}(F_Y^{-1}(\Phi_1(Z_0))|x_0, \boldsymbol{x}, \boldsymbol{\beta}, \boldsymbol{\theta}, \boldsymbol{\gamma}) = \int F_Y^{-1}(\Phi_1(z_0)) p(z_0|x_0, \boldsymbol{x}, \boldsymbol{\beta}, \boldsymbol{\theta}, \boldsymbol{\gamma}) \mathrm{d}z_0$$

$$= \int F_Y^{-1}(\Phi_1(z_0)) \frac{1}{s_0} \phi_1\left((z_0 - s_0 \boldsymbol{b}_0'\boldsymbol{\beta})/s_0\right) \mathrm{d}z_0, \tag{10}$$

where $\boldsymbol{b}_0$ is the vector of basis terms evaluated at the covariate value $x_0$, and $s_0 = [1 + \boldsymbol{b}_0' P(\boldsymbol{\theta})^{-1} \boldsymbol{b}_0]^{-1/2}$. We employ $\hat{F}_Y$ for the marginal distribution function of $Y_0|x_0$, and compute the integral above using standard univariate numerical methods. Finally, the estimator

$$\mathbb{E}(Y_0|x_0, \boldsymbol{x}, \boldsymbol{y}) \approx \frac{1}{J} \sum_{j=1}^{J} \mathbb{E}\left(Y_0|x_0, \boldsymbol{x}, \boldsymbol{\beta}^{[j]}, \boldsymbol{\theta}^{[j]}, \boldsymbol{\gamma}^{[j]}\right) = \hat{f}(x_0) \tag{11}$$

can be computed from the output $\{\boldsymbol{\beta}^{[j]}, \boldsymbol{\theta}^{[j]}, \boldsymbol{\gamma}^{[j]}; j = 1, \ldots, J\}$ of the MCMC scheme. It can also be useful to estimate $m(x_0) \equiv \mathbb{E}(Z_0|x_0)$ at (5), for which we use the posterior predictive mean

$$\mathbb{E}(Z_0|x_0, \boldsymbol{x}, \boldsymbol{y}) = \int \mathbb{E}(Z_0|x_0, \boldsymbol{x}, \boldsymbol{\beta}, \boldsymbol{\theta}, \boldsymbol{\gamma}) p(\boldsymbol{\beta}, \boldsymbol{\theta}, \boldsymbol{\gamma}|\boldsymbol{x}, \boldsymbol{y}) \mathrm{d}(\boldsymbol{\beta}, \boldsymbol{\theta}, \boldsymbol{\gamma})$$

$$= \int s_0 \boldsymbol{b}_0' \boldsymbol{\beta}\, p(\boldsymbol{\beta}, \boldsymbol{\theta}, \boldsymbol{\gamma}|\boldsymbol{x}, \boldsymbol{y}) \mathrm{d}(\boldsymbol{\beta}, \boldsymbol{\theta}, \boldsymbol{\gamma}) \approx \boldsymbol{b}_0'\left(\frac{1}{J}\sum_{j=1}^{J} s_0^{[j]} \boldsymbol{\beta}^{[j]}\right) = \hat{m}(x_0),$$

where $s_0^{[j]} = [1 + \boldsymbol{b}_0' P(\boldsymbol{\theta}^{[j]})^{-1} \boldsymbol{b}_0]^{-1/2}$.

For the BVSC, the vector $\boldsymbol{\beta}$ is not generated as part of the Gibbs sampler in Section 3.2.3. Therefore, to compute this function estimator, it is necessary to generate from the Gaussian distribution $\boldsymbol{\beta}_\gamma^{[j]} \sim \boldsymbol{\beta}_\gamma|\boldsymbol{x}, \boldsymbol{\gamma}, \boldsymbol{y}$ at the end of each sweep, and set the remaining elements of $\boldsymbol{\beta}^{[j]}$ to zero. Also, note that for this case $s_0$ is a function of all covariate values $(\boldsymbol{x}', x_0')'$, whereas for the HSC and PSC the standardizing constant $s_0$ is a function of $x_0$ only.

We compute the function estimators $\hat{f}$ and $\hat{m}$ over a grid of values for $x_0$. Note that $f^{[j]}(x_0) = \mathbb{E}(Y_0|x_0, \boldsymbol{x}, \boldsymbol{\beta}^{[j]}, \boldsymbol{\theta}^{[j]}, \boldsymbol{\gamma}^{[j]})$ and $m^{[j]}(x_0) = s_0^{[j]} \boldsymbol{b}_0'\boldsymbol{\beta}^{[j]}$ are draws from the posterior distribution of each function at point $x_0$. Therefore, posterior $(100-\alpha)\%$ probability intervals can be computed for $f$ and $m$ at point $x_0$ by ordering these draws and counting off $\alpha/2\%$ of the highest and lowest values in the standard Bayesian fashion.

Evaluation of $\hat{f}(x_0)$ requires $J$ numerical integrations for each value of $x_0$. An alternative estimator that is faster to compute, is to plug in the point estimators for the quantities in (10), giving $\tilde{f}(x) = \int F_Y^{-1}(\Phi_1(z_0)) \frac{1}{\hat{s}_0} \phi_1\left((z_0 - \hat{m}(x_0))/\hat{s}_0\right) \mathrm{d}z_0$ with $\hat{s}_0 = \frac{1}{J}\sum_{j=1}^{J} s_0^{[j]}$. This involves computing only a single univariate numerical integral. Tab. 2 summarizes the functional relationships in the copula model, the Bayesian posterior means and their MCMC estimators.

## 3.4 Predictive Densities

The predictive density $p(y_0|x_0)$ of a new observation of the response $Y_0$, given a new covariate value $x_0$, is estimated using its posterior predictive density

$$p(y_0|x_0, \boldsymbol{x}, \boldsymbol{y}) = p(y_0|x_0, \boldsymbol{x}, \boldsymbol{\beta}, \boldsymbol{\theta}, \boldsymbol{\gamma}) p(\boldsymbol{\beta}, \boldsymbol{\theta}, \boldsymbol{\gamma}|\boldsymbol{x}, \boldsymbol{y}).$$

If $z_0 = \Phi_1^{-1}(F_Y(y_0))$, then $\left|\frac{dz_0}{dy_0}\right| = \frac{p_Y(y_0)}{\phi_1(z_0)}$, and by changing variables from $y_0$ to $z_0$,

$$\begin{aligned}
p(y_0|x_0, \boldsymbol{x}, \boldsymbol{\beta}, \boldsymbol{\theta}, \boldsymbol{\gamma}) &= \frac{p_Y(y_0)}{\phi_1(z_0)} p(z_0|x_0, \boldsymbol{x}, \boldsymbol{\beta}, \boldsymbol{\theta}, \boldsymbol{\gamma}) \\
&= \frac{p_Y(y_0)}{\phi_1\left(\Phi_1^{-1}(F_Y(y_0))\right)} \frac{1}{s_0} \phi_1 \left( \frac{\Phi_1^{-1}(F_Y(y_0)) - m(x_0)}{s_0} \right),
\end{aligned}$$

which follows from (5). We estimate the expression above using the estimate of the marginal density $\hat{p}_Y$, and the Monte Carlo iterates as

$$\hat{p}(y_0|x_0) = \frac{\hat{p}_Y(y_0)}{\phi_1(\Phi_1^{-1}(\hat{F}_Y(y_0)))} \left\{ \frac{1}{J} \sum_{j=1}^{J} \frac{1}{s_0^{[j]}} \phi_1 \left( \frac{\Phi_1^{-1}(\hat{F}_Y(y_0)) - m^{[j]}(x_0)}{s_0^{[j]}} \right) \right\}. \quad (12)$$

It is also possible to estimate the predictive density $p(z_0|x_0)$ of $Z_0$ given $x_0$ on the latent space, using the posterior predictive density

$$\begin{aligned}
p(z_0|x_0, \boldsymbol{x}, \boldsymbol{y}) &= \int p(z_0|x_0, \boldsymbol{x}, \boldsymbol{\beta}, \boldsymbol{\theta}, \boldsymbol{\gamma}) p(\boldsymbol{\beta}, \boldsymbol{\theta}, \boldsymbol{\gamma}|\boldsymbol{x}, \boldsymbol{y}) \mathrm{d}(\boldsymbol{\beta}, \boldsymbol{\theta}, \boldsymbol{\gamma}) \\
&= \int \frac{1}{s_0} \phi_1 \left( \frac{z_0 - s_0 \boldsymbol{b}_0' \boldsymbol{\beta}}{s_0} \right) p(\boldsymbol{\beta}, \boldsymbol{\theta}, \boldsymbol{\gamma}|\boldsymbol{x}, \boldsymbol{y}) \mathrm{d}(\boldsymbol{\beta}, \boldsymbol{\theta}, \boldsymbol{\gamma}),
\end{aligned}$$

which is estimated readily using the Monte Carlo iterates as $\hat{p}(z_0|x_0) = \frac{1}{J} \sum_{j=1}^{J} \frac{1}{s_0^{[j]}} \phi_1((z_0 - m^{[j]}(x_0))/s_0^{[j]})$.

## 3.5 Illustration

To illustrate the posterior dependence structure and function estimates of each copula, we simulate $y_i \sim N(h_3(x_i), 0.5^2)$ using the same covariate values as employed previously in Section 2.3 and function $h_3$ defined in Section 4 below. Fig. 1(b) gives a scatterplot of the resulting data, along with a plot of $h_3$. We estimate the marginal of the data using the adaptive kernel density estimator, and compute $p(\boldsymbol{\theta}, \boldsymbol{\gamma}|\boldsymbol{x}, \boldsymbol{y})$ for the same four copula/basis combinations employed previously.

Using the draws from the posteriors, we compute the surface of Spearman correlations $\rho_\pi^s$ and plot these on the right-hand side of Fig. 3 for comparison with those evaluated previously using draws from the priors $\pi_0$. The general features of the prior dependence structures discussed in Section 2.3 transfer to the posteriors, although there are some notable differences, and we make four observations. First, the posterior dependence structure of the PSC/B-spline in panel (b) is sharper than its prior in panel (a). Second, the posterior dependence structure of the HSC/B-spline in panel (d) is asymmetric along the line $x_{0,1} = x_{0,2}$, with higher smoothing for covariate values around 0.2, 0.3 and close to 1. This local adaptivity is evident in the posterior, but not the prior. Third, when the HSC is combined with the augmented Fourier basis in panel (f), smoothing is non-monotonic in $|x_{0,1} - x_{0,2}|$ because the basis terms are also. Last, the BVSC with a regression spline basis in panel (h) has a posterior level of smoothing that is higher than that of the prior in panel (g). Yet the level of smoothing varies greatly with the value of the covariate, with more smoothing for values greater than 0.5, and less for values around 0.3.

Fig. 1(c,d) plots posterior function estimates for this data for each of the four copula/basis combinations. The estimator $\hat{f}$ was used for the PSC and HSC, and $\tilde{f}$ for the BVSC. All function estimates track the data well, although those from the PSC and HSC models under-smooth on the right-hand side of the function. In contrast, the BVSC produces a smoother estimate, which is because Bayesian variable selection is known to be a highly locally adaptive regularization method. To compute 1,000 sweeps of the MCMC schemes it took approximately 13, 27 and 3.5 seconds for the HSC, PSC and BVSC, respectively, when implemented in serial using Matlab on a standard desktop.

# 4 Univariate Simulation

To illustrate the effectiveness of the copula smoother we undertake a simulation study. The PSC with a B-spline basis and nonparametric margins $\hat{F}_Y$ is compared to a Bayesian P-spline with the same basis and Gaussian disturbances (labeled as PS).

## 4.1 Simulation Design

We consider the following three univariate test functions:

$$h_1(x) = 2x-1, \quad h_2(x) = \sin(10\pi x), \quad h_3(x) = \frac{1}{4}\left[\frac{1}{0.05}\phi_1((x-0.15)/0.05) + \frac{1}{0.2}\phi_1((x-0.6)/0.2)\right].$$

For each function $j = 1, 2, 3$, we generate $n = 100$ observations from three distributions:

Case 1, Normal: $\quad Y_{1j} = h_j(x) + \varepsilon_1, \quad\quad\quad\quad\quad\quad\quad \varepsilon_1 \sim \text{iid}\, N(0, 0.5^2)$

Case 2, Log-normal: $\quad Y_{2j} = h_j(x) + \varepsilon_2 - \mathbb{E}(\varepsilon_2), \quad\quad\quad \varepsilon_2 \sim \text{iid}\, \text{LN}(-2.89, 1.5^2)$

Case 3, Implicit Copula: $\quad Y_{3j} = F_{\text{Gam}}^{-1}(z_j; 3, 2),\, z_j = h_j(x) + \varepsilon_3, \quad \varepsilon_3 \sim \text{iid}\, N(0, r_j^2).$

where $F_{\text{Gam}}$ is a Gamma distribution function, and LN the lognormal distribution. The distribution of $Y_{lj}$, $l = 1, 2, 3$ is defined conditional on the covariate $x$, which we generate independently from a uniform distribution on $(0, 1)$. Note that the distribution in Case 1 matches that of the Gaussian P-spline, while that in Case 3 matches that of the implicit copula model with a Gamma margin. The distribution in Case 2 matches neither model.

The true regression and noise functions are $f_j(x) \equiv \mathbb{E}(Y_{lj}|x)$ and $v_{lj}(x) \equiv \text{Var}(Y_{lj}|x)$, and in each case the signal-to-noise ratio is $\text{SNR}_{lj} \equiv \text{range}(f_j(x))/(v_{lj}(x))^{1/2} = 4$ over domain of the covariate $0 \leq x \leq 1$. In Cases 1 and 2 ($l = 1, 2$), it is straightforward to see that $f_j = h_j$, $v_{lj}(x) = \text{Var}(\varepsilon_j)$ is a constant and that $\text{SNR}_{lj} = 4$. However, in Case 3, $f_j$ and $v_{lj}$ are more complex functions of $h_j$, with

$$f_j(x) = \int y_j p(y_j|x) dy_j = \int F_{\text{Gam}}^{-1}(\Phi_1(z_j); 3, 2) \frac{1}{r_j} \phi_1((z_j - h_j(x))/r_j) \, dz_j,$$

$$v_{lj}(x) = \mathbb{E}(Y_j^2|x) - f_j(x)^2 = \int [F_{\text{Gam}}^{-1}(\Phi_1(z_j); 3, 2)]^2 \frac{1}{r_j} \phi_1((z_j - h_j(x))/r_j) \, dz_j - f_j(x)^2,$$

where the integrals are computed numerically. Setting $\text{SNR}_{3j} = 4$ over $0 \leq x \leq 1$, it is possible to solve for $r_j$ to get $r_1 = 0.48$, $r_2 = 0.47$ and $r_3 = 0.58$ for the three functions. For each of the nine combinations of Case $l$ and function $h_j$ we simulated 100 replicates, leading to a total of 900 datasets.

For both the PSC and the PS the same cubic B-spline basis is employed with equally spaced knots and $\dim(\boldsymbol{\beta}) = 32$. As as outlined in Section 3.2.1, the precision matrix of an AR(1) is used for constructing the PSC implicit copula. For the PS the popular first order random walk prior (Lang and Brezger, 2004) is used, although the results are almost identical when the precision matrix of an AR(1) model is employed.

## 4.2 Measures of Performance

We consider three measures of the quality of the fitted statistical models. The first is a measure of the accuracy of the point estimate of the regression function, and is the root mean square error $\text{RMSE}(f, \hat{f}) = \left(\frac{1}{n}\sum_{i=1}^{n}(\hat{f}(x_i) - f(x_i))^2\right)^{1/2}$ computed over the data points. For the PSC model the regression function estimator is given at (11), whereas for the PS it is $\hat{f}(x_i) = \boldsymbol{b}_i' \mathbb{E}(\boldsymbol{\beta}|\boldsymbol{y})$, which we compute using the BayesX software (Belitz et al., 2015).

The second measure is based on the Kullback-Leibler Divergence (KLD) between the density $p(y|x)$ of the data generating process, and its estimate $\hat{p}(y|x)$, given by

$$\text{KLD}_x(p||\hat{p}) = \int p(y|x) \log\left(\frac{p(y|x)}{\hat{p}(y|x)}\right) dy.$$

To compute the KLD, note that for Cases 1 and 2 the density $p(y|x)$ is a normal and log-normal distribution, respectively. For Case 3, the density is

$$p(y|x) = \frac{p_{\text{Gam}}(y;3,2)}{\phi_1(\Phi_1^{-1}(F_{\text{Gam}}(y;3,2)))r_j}\phi_1\left(\frac{\Phi_1^{-1}(F_{\text{Gam}}(y;3,2)) - h_j(x)}{r_j}\right),$$

where $p_{\text{Gam}}$ is a Gamma density function.

For the PSC, the density estimator is given at (12). For the regular PS, $\hat{p}(y_0|x_0) = (1/\hat{\sigma})\phi_1((y_0 - \hat{f}(x_0))/\hat{\sigma})$, with point estimators $\hat{\sigma}$ and $\hat{f}$. The integral can be computed analytically for the Case 1/PS combination and numerically for the other five combinations of estimator and Case; see the Online Appendix Tab. A. Finally, we report the mean KLD over an equally-spaced partition $0 = \tilde{x}_1 < \ldots < \tilde{x}_N = 1$ of the covariate, giving $\text{MKLD}(p||\hat{p}) = \frac{1}{N}\sum_{i=1}^{N}\text{KLD}_{\tilde{x}_i}(p||\hat{p})$, where we set $N = 100$. This metric measures the accuracy of $\hat{p}(\cdot|x_0)$.

The third and final measure is of predictive performance. This is the mean logarithmic score computed by ten-fold cross-validation. For a given dataset, we compute this by partitioning the data into ten sub-samples, denoted as $\{(y_{i,k}, x_{i,k}); i = 1, \ldots, n_k\}$ for $k = 1, \ldots, 10$. For sub-sample $k$, we compute the density estimator using the remaining 9 sub-samples as the training data, and denote these as $\hat{p}_k(y|x)$. The ten-fold mean logarithmic score is then $\text{MLS} = \frac{1}{10}\sum_{k=1}^{10}\frac{1}{n_k}\sum_{i=1}^{n_k}\log \hat{p}_k(y_{i,k}|x_{i,k})$. Here $n = 100$, so that we set $n_k = 10$, giving sub-samples of equal size.

## 4.3 Results

Fig. 4 compares the accuracy of the PSC and PS estimators of the regression functions using the RMSE metric. There are nine panels: one for each combination of Cases 1,2,3 and test functions $h_1, h_2, h_3$. The accuracy of the two function estimators is similar, even in Case 1 where the PS estimator is the correct model. This is reassuring because the Bayesian P-spline is known to be a highly competitive regression function estimator (Lang and Brezger, 2004; Scheipl et al., 2012). To illustrate, Fig. A of the Online Appendix plots the true regression function $f_j$ and both estimates for a single replicate of data in each case, along with a scatterplot of the data. The function estimates are similar and track the data well. However, the PSC and PS density estimators differ substantially. Fig. 5 presents boxplots of the MKLD metric for each of the nine combinations. The PS is slightly more accurate than the PSC in Case 1, which is unsurprising because the PS is a conditionally Gaussian model and matches the data generating process. But in the two non-Gaussian cases — including Case 2 where neither model is correct — the PSC density estimator is substantially more accurate. The same conclusions are drawn from Fig. B of the Online Appendix, which presents equivalent boxplots for the MLS metric. Thus, using the copula model also results in a substantial increase in the accuracy of the predictive distributions for non-Gaussian data.

## 5 Multivariate Extensions

The implicit copula (and the resulting copula smoother) can be extended to account for multiple covariates in two ways. The first way is by constructing the implicit copula of an additive model, while the second is by employing a radial basis. We explain how to do so below, and illustrate using two real data examples.

### 5.1 Additive Copula Smoother

#### 5.1.1 Implicit Copula

Consider replacing (1) with the additive regression

$$\tilde{Z}_i = \sum_{l=1}^{L} \tilde{m}_l(x_{il}) + \varepsilon_i, \quad \text{for } i = 1, \ldots, n. \tag{13}$$

As before, each function is modeled as a linear combination of basis functions $\tilde{m}_l(x_l) = \sum_j^{p_l} b_{lj}(x_l)\beta_{lj}$, with corresponding design matrix $B_l$ and coefficient vector $\boldsymbol{\beta}_l = (\beta_{l1},\ldots,\beta_{lp_l})'$. Then the additive regression can be written as the linear model at (2), but where $B = [B_1|\cdots|B_L]$ is an $(n \times \sum_{l=1}^L p_l)$ concatenated design matrix and $\boldsymbol{\beta}' = (\boldsymbol{\beta}'_1,\ldots,\boldsymbol{\beta}'_L)$. Our objective here is to construct the implicit copula of this additive model.

A global intercept parameter is not included in (13) because it is unidentified in its implicit copula. To ensure identifiability of $\boldsymbol{\beta}$, we centre all but one $\tilde{m}_l$ around zero, so that $\mathbf{1}'\tilde{m}_l(\boldsymbol{x}_l) = \mathbf{1}'B_l\boldsymbol{\beta}_l = 0$, for $l = 1,\ldots,L-1$, with $\mathbf{1}$ an $n$-vector of ones. To regularize each vector $\boldsymbol{\beta}_l$, we assume the same shrinkage prior at (3), but with these constraints, so that

$$p(\boldsymbol{\beta}_l|\boldsymbol{x},\boldsymbol{\theta}_l,\boldsymbol{\gamma}_l) \propto \begin{cases} \phi_{p_l}(\boldsymbol{\beta}_l;\mathbf{0},P(\boldsymbol{\theta})^{-1})I(\mathbf{1}'B_l\boldsymbol{\beta}_l = 0) & \text{if } l = 1,\ldots,L-1 \\ \phi_{p_l}(\boldsymbol{\beta}_l;\mathbf{0},P(\boldsymbol{\theta})^{-1}) & \text{if } l = L \end{cases},$$

where each prior is strictly proper. Setting $\boldsymbol{x}_l = (x_{1l},\ldots,x_{nl})'$ and $P(\boldsymbol{\theta}) = \text{bdiag}(P(\boldsymbol{\theta}_1),\ldots,P(\boldsymbol{\theta}_L))$ as a block diagonal matrix, $\boldsymbol{\beta}$ can be integrated out as a linearly constrained normal, giving

$$\tilde{\boldsymbol{Z}}|\boldsymbol{x}_1,\ldots,\boldsymbol{x}_L,\boldsymbol{\theta},\boldsymbol{\gamma} \sim N(\mathbf{0},(I + BP(\boldsymbol{\theta})^{-1}B')),$$

as in Section 2.1. Standardization of $\tilde{\boldsymbol{Z}}$ and formation of the implicit copula then proceeds as in the univariate case, but where $\boldsymbol{b}_i = (\boldsymbol{b}'_{i1},\ldots,\boldsymbol{b}'_{iL})'$, $\boldsymbol{b}_{il} = (b_{l1}(x_{il}),\ldots,b_{lp_l}(x_{il}))'$,

$$s_i = (1 + \boldsymbol{b}'_i P(\boldsymbol{\theta})^{-1}\boldsymbol{b}_i)^{-1/2} = \left(1 + \sum_{l=1}^L \boldsymbol{b}'_{il} P(\boldsymbol{\theta}_l)^{-1}\boldsymbol{b}_{il}\right)^{-1/2}, \text{ and}$$

$$\Omega^{-1} = \text{bdiag}\left(B'_1 B_1 + P(\boldsymbol{\theta}_1),\ldots,B'_L B_L + P(\boldsymbol{\theta}_L)\right).$$

The posterior can be evaluated using the MCMC schemes outlined in the univariate case, with one change. When generating $\boldsymbol{\beta}$, we generate each sub-vector $\boldsymbol{\beta}_l$ conditional on the other elements of $\boldsymbol{\beta}$. For $l = 1,\ldots,L-1$ this involves generating from a constrained normal using the fast algorithm in Rue and Held (2005, Alg. 2.6). Further details on how to implement the MCMC scheme for the PSC are given in Appendix Part C.

### 5.1.2 Function Estimation and Partial Residuals

For a new observation $(Y_0, x_{01},\ldots,x_{0L})$ on the response and covariates, the regression surface is $f(x_{01},\ldots,x_{0L}) \equiv E(Y_0|x_{01},\ldots,x_{0L})$. It can be estimated in the same manner as in

Section 3.3, but where

$$m(x_{01},\ldots,x_{0L}) = s_0 \bm{b}'_0 \bm{\beta} = s_0 \sum_{l=1}^{L} \bm{b}'_{0l} \bm{\beta}_l = \sum_{l=1}^{L} m_l(x_{0l}),$$

with $s_0$ as defined above and $m_l(x_{0l}) = s_0 \bm{b}'_{0l} \bm{\beta}_l$.

Even though the relationship at (13) is additive in the covariates, the regression surface $f$ is not. This means that partial residuals — a popular diagnostic for additive models (Hastie and Tibshirani, 1990) — cannot be easily defined for $\bm{y}$. However, they can be for the values $z_1, \ldots, z_n$ as follows.

**Definition 2.** *For the $i$-th observation and $j$-th effect of the additive basis copula, $i = 1, \ldots, n$ and $j = 1, \ldots, L$, we define the $j$-th partial residual $\epsilon_{i,j}$ as*

$$\epsilon_{i,j} = z_i - \sum_{l \neq j} m_l(x_i) = z_i - s_i \sum_{l \neq j} \bm{b}'_{il} \bm{\beta}_l,$$

*where $s_i$ is defined above.*

If the model is correct, then from (5), the partial residual $\epsilon_{i,j}$ is a realization from a $N(m_j(x_i), s_i)$ distribution.

### 5.1.3 Example: Boston Housing Data

To illustrate we employ the Boston housing data (Harrison and Rubinfeld, 1978). The data comprise observations on the median value (PRICE) of residential homes in $n = 506$ Boston census tracts. Also recorded are five continuous hedonic variables (NOX, RM, DIS, LSTAT and TAX). The dataset is a common test for flexible regression methods with PRICE as the response. Fig. 7 plots the histogram of PRICE, which is far from Gaussian. Regression models with normal disturbances produce poor estimates of the functional relationships; for example, in their analysis Smith and Kohn (1996) estimate a Box-Cox transformation of the response PRICE and model the errors as a mixture of two normals.

We model PRICE using the additive PSC smoother with the five continuous variables as covariates. Fig. 7 plots an adaptive kernel density estimate, from which the marginal distribution function $\hat{F}_Y$ is computed. For each covariate, a cubic B-spline basis with equally spaced knots and where $\dim(\bm{\beta}_l) = 22$ was employed. Fig. 6 presents summaries of the functional relationships from the fitted copula smoother. The left-hand panels (a,c,e,g,i) plot

'slices' of $\hat{f}$ against each of the five covariates, where in each panel the values of the other four covariates are fixed to those of the observation with the median PRICE. Also plotted are the equivalent slices of the 95% posterior probability interval for $f$. For comparison, we estimate an additive P-spline with the same basis using the BayesX software. Panels (a,c,e,g,i) depict the equivalent function estimates from this additive model, and they differ from those of the copula model. The right-hand panels (b,d,f,h,j) show the posterior mean of $m_l(x_{0l}) = s_0 \boldsymbol{b}'_{0l}\boldsymbol{\beta}_l$, along with 95% posterior probability intervals for $m_l(x_{0l})$, for $l = 1, \ldots, 5$. The scatterplots are of the partial residuals $\{\epsilon_{1,l}, \ldots, \epsilon_{n,l}\}$.

To compare the two models, we compute the mean logarithmic score for a ten-fold cross validation as in Section 4. For the copula model MLS $= -2.47$, compared to MLS $= -2.86$ for the additive P-spline, suggesting that the copula model has more accurate predictive densities. To highlight why this is the case, Fig. 8 plots the predictive densities $\hat{p}(y_0|\boldsymbol{x}_0)$ from both fitted models for six representative observations. These are the observations at quantiles $q = 0.025, 0.2, 0.4, 0.6, 0.8, 0.975$ of the PRICE distribution. The predictive distributions from the copula model are generally tighter (i.e. more 'sharp'), and feature a high degree of asymmetry throughout. The predictive density in panel (f) has a spike at PRICE=$50,000, which is caused by a few high-valued observations that are unexplained by the covariates. Earlier analysis (Smith and Kohn, 1996) treats these as outliers, but in the copula model they are captured by the estimated marginal $\hat{F}_Y$ in Fig. 7. In contrast, these outliers are not well captured using the P-spline, which has a necessarily Gaussian predictive density in panel (f).

## 5.2 Radial Bases

Another approach to account for multiple covariates is to employ a radial basis (Powell, 1987). In this case, at (1) the function $\tilde{m}(x_{i1}, \ldots, x_{iL}) = \sum_{j=1}^{p} \beta_j b_j(x_{i1}, \ldots, x_{iL})$, where $b_j(x_1, \ldots, x_L) = \zeta(\|(x_1, \ldots, x_L) - (k_{j1}, \ldots, k_{jL})\|)$ is a radial basis function, $\|\cdot\|$ is the Euclidean distance, and $(k_{j1}, \ldots, k_{jL})$ is a multivariate knot. The $p$ knots are typically a subsample of observed covariate values, and here we select a random subsample of $p = 100$ values inside their convex hull. If the covariates are scaled to the unit cube, typical choices for $\zeta$ are a Gaussian kernel $\zeta(x) = \frac{1}{8}\phi_1(x/8)$ and a thin-plate spline $\zeta(x) = x^2 \ln(x)$ (Bookstein, 1989). Regularization using the autoregressive prior in Section 2.2.1 is inappropriate because the radial basis terms do not have an adjacent ordering. Therefore, for radial bases we only consider copula smoothers with the HSC and BVSC implicit copulas.

To illustrate, we model the logarithm of prices ($Y$) of $n = 11,375$ fine art prints sold at international auctions during 2015. These are the 'realized prices', which include auction fees and taxes, and are converted into U.S. dollars at the exchange rate on the date of sale. Fig. 9(a) plots the histogram of $Y$, showing that even after the logarithmic transformation, prices are far from Gaussian. Also plotted is the kernel density estimate, from which we compute $z_i = \hat{F}_Y(y_i)$ for $i = 1, \ldots, n$. We consider two covariates: the logarithm of the area of the print ($X_1$), and the logarithm of the edition size ($X_2$). The data were sourced from MutualArt, which is a leading art investment fund. In general, prints with lower area and from larger editions are likely to be worth less (Pesando and Shum, 2008).

We standardize the covariates to the unit square and fit copula smoothers with both the HSC and BVSC, and Gaussian kernel and thin-plate spline radial bases. Despite the very high dimension of the copula, 1,000 sweeps of the MCMC scheme takes approximately 11 mins for the HSC, and 17 mins for the BVSC; both implemented in Matlab and run in serial on a standard desktop. Fig. 9(b–e) plots the bivariate function estimates $\hat{f}$ from the four fitted models. Throughout, prints with larger sizes (i.e. higher values of $X_1$) tend to be worth more, while the impact of the edition size is more mixed. The surfaces in panels (c,d,e) are similar, while that in panel (b) is smoother, so that the HSC with a thin-plate spline basis exhibits a higher level of regularization.

# 6 Discussion

The paper presents a general approach to construct the implicit copula of regularized regression smoothers with Gaussian disturbances. Three diverse shrinkage priors are considered in detail, although the approach can also be employed with other conjugate priors. A Gaussian copula is first constructed by integrating out $\boldsymbol{\beta}$, but conditioning on the the hyper-parameters $(\boldsymbol{\theta}, \boldsymbol{\gamma})$. The implicit copula is then formed by mixing over their prior or posterior distribu-tions. This conditioning trick greatly simplifies the computation of the implicit copula, which is much harder to compute via inversion of the distribution $\tilde{\bm{Z}}|\bm{x}$ directly. We stress here that the implicit copula is not a Gaussian copula, and can have a very different dependence structure as illustrated in Fig. 3.

The implicit copulas provide a convenient way to compare the smoothing properties of the different shrinkage priors and bases. They also can be used to extend the regularized regression smoothers to non-Gaussian data by combining them with flexible margins. In this case, the copula is of $n$ dimensions. Nevertheless, the proposed MCMC schemes can be used to compute the posterior function estimates efficiently, even for higher sample sizes such as in the print price example with $n = 11,375$.

We finish by mentioning promising directions for extension of our proposed approach. First, the implicit copulas for other popular conjugate priors for regularization (Liang et al., 2008; Scheipl et al., 2012) may be constructed. Second, regression smoothers with elliptical error distributions beyond the Gaussian can be considered. When combined with conjugate priors, application of the conditioning trick will result in the implicit copula being a mixture over the corresponding elliptical copula. Third, while we use the copula smoother to model non-Gaussian continuous data, the copula can also be employed for modeling discrete-valued or mixed data. For these cases, new ways to simulate iterates from the posterior distribution of the hyper-parameters are required, such as data augmentation (Pitt et al., 2006) and pseudo-marginal MCMC (Gunawan et al., 2016).

# Appendix A    Proof of Corollary 1

Following Definition (2.10.6., Nelson, 2006), it is sufficient to show:

1. For every $\bm{u} \in [0,1]^n$, if at least one coordinate of $\bm{u}$ is zero then
$$C_\pi(\bm{u}|\bm{x}) = \int\int C(\bm{u}|\bm{x},\bm{\theta},\bm{\gamma})\pi(\bm{\theta},\bm{\gamma})\mathrm{d}(\bm{\theta},\bm{\gamma}) = 0\int\int \pi(\bm{\theta},\bm{\gamma})\mathrm{d}(\bm{\theta},\bm{\gamma}) = 0\,.$$
which follows because $C(\bm{u}|\bm{x},\bm{\theta}\bm{\gamma})$ is a copula function. Similarly, if all coordinates of $\bm{u}$ are 1 except $u_k$, then
$$C_\pi(\bm{u}|\bm{x}) = \int\int C(\bm{u}|\bm{x},\bm{\theta},\bm{\gamma})\pi(\bm{\theta},\bm{\gamma})\mathrm{d}(\bm{\theta},\bm{\gamma}) = x_k\int\int \pi(\bm{\theta},\bm{\gamma})\mathrm{d}(\bm{\theta},\bm{\gamma}) = x_k.$$

2. For every $\boldsymbol{a}, \boldsymbol{b} \in [0,1]^n$ such that the $C_\pi(\cdot|\boldsymbol{x})$-volume $\boldsymbol{a} \leq \boldsymbol{b}$, then $V_{C_\pi}([\boldsymbol{a}, \boldsymbol{b}]) \geq 0$ since the Gaussian copula and the priors are proper densities. We refer to (2.10.1., Nelson, 2006) for the definition of the $C$-volume of $[\boldsymbol{a}, \boldsymbol{b}]$.

# Appendix B   Implementation of Sampling Schemes

In this appendix we provide the derivatives and computational details required for the efficient implementation of the three sampling schemes in Section 3.2.

## B.1   P-Spline Implicit Copula

The derivatives for the proposal densities of $\upsilon = \log(\tau^2)$ are:

$$\frac{\partial l_\upsilon}{\partial \upsilon} = -\frac{1}{2}\left(\dim(P(\psi))-1\right) + \frac{1}{2\exp(\upsilon)}\boldsymbol{\beta}'P(\psi)\boldsymbol{\beta} - \left(\frac{\exp(\upsilon)}{4b_{\tau^2}}\right)^{\frac{1}{2}}$$
$$- \frac{1}{2}\sum_{i=1}^{n}\frac{1}{s_i^2}\frac{\partial}{\partial \upsilon}s_i^2 - \frac{1}{2}\left(\boldsymbol{z}'\left[\frac{\partial}{\partial \upsilon}(SS')^{-1}\right]\boldsymbol{z} - 2\boldsymbol{\beta}'B'\left[\frac{\partial}{\partial \upsilon}S^{-1}\right]\boldsymbol{z}\right)$$

$$\frac{\partial^2 l_\upsilon}{\partial \upsilon^2} = -\frac{1}{2\exp(\upsilon)}\boldsymbol{\beta}'P(\psi)\boldsymbol{\beta} - \left(\frac{\exp(\upsilon)}{16b_0}\right)^{\frac{1}{2}} - \frac{1}{2}\sum_{i=1}^{n}\left(\left[\frac{\partial}{\partial \upsilon}\frac{1}{s_i^2}\right]\frac{\partial}{\partial \upsilon}s_i^2 + \frac{1}{s_i^2}\frac{\partial^2}{\partial \upsilon^2}s_i^2\right)$$
$$- \frac{1}{2}\left(\boldsymbol{z}'\left[\frac{\partial^2}{\partial \upsilon^2}(SS')^{-1}\right]\boldsymbol{z} - 2\boldsymbol{\beta}'B'\left[\frac{\partial^2}{\partial \upsilon^2}S^{-1}\right]\boldsymbol{z}\right).$$

The derivatives for the proposal densities of $\xi = \log\left(\frac{\psi-\epsilon}{1-\epsilon-\psi}\right)$ are:

$$\frac{\partial l_\xi}{\partial \xi} = 1 - \frac{2\exp(\xi)}{1+\exp(\xi)} + \frac{1}{2}\frac{\partial}{\partial \xi}\log(\det(\Delta(g^{-1}(\xi)))) - \frac{\partial}{\partial \xi}\frac{\boldsymbol{\beta}'P(g^{-1}(\xi))\boldsymbol{\beta}}{2\tau^2}$$
$$- \frac{1}{2}\sum_{i=1}^{n}\frac{\frac{\partial}{\partial \xi}s_i^2}{s_i^2} - \frac{1}{2}\left(\boldsymbol{z}'\frac{\partial}{\partial \xi}(SS')^{-1}\boldsymbol{z} - 2\boldsymbol{\beta}'B'\frac{\partial}{\partial \xi}S^{-1}\boldsymbol{z}\right)$$

$$\frac{\partial^2 l_\xi}{\partial \xi^2} = -\frac{2\exp(\xi)}{(1+\exp(\xi))^2} + \frac{1}{2}\frac{\partial^2}{\partial \xi^2}\log(\det(\Delta(g^{-1}(\xi)))) - \frac{\partial^2}{\partial \xi^2}\frac{\boldsymbol{\beta}'P(g^{-1}(\xi))\boldsymbol{\beta}}{2\tau^2}$$
$$- \frac{1}{2}\sum_{i=1}^{n}\frac{\partial^2}{\partial \xi^2}\frac{\frac{\partial}{\partial \xi}s_i^2}{s_i^2} - \frac{1}{2}\left(\boldsymbol{z}'\frac{\partial^2}{\partial \xi^2}(SS')^{-1}\boldsymbol{z} - 2\boldsymbol{\beta}'B'\frac{\partial^2}{\partial \xi^2}S^{-1}\boldsymbol{z}\right).$$

The derivation of these derivatives, including the matrix derivatives of $P$, can be found in the Online Appendix Part A.

## B.2 Horseshoe Implicit Copula

The derivatives for the proposal densities of $\upsilon = \log(\lambda_j^2)$ are

$$\frac{\partial}{\partial \log(\lambda_j^2)} \log(p(\log(\lambda_j^2)|\boldsymbol{\beta}, \tau, \boldsymbol{y}, \boldsymbol{x})) = -\frac{1}{2}\sum_{i=1}^{n} \frac{\frac{\partial}{\partial \log(\lambda_j^2)} s_i^2}{s_i^2} - \frac{1}{2}\boldsymbol{z}'(\frac{\partial}{\partial \log(\lambda_j^2)} SS')^{-1}\boldsymbol{z}$$

$$+\boldsymbol{\beta}'B'\frac{\partial}{\partial \log(\lambda_j^2)} S^{-1}\boldsymbol{z} - \frac{1}{2}\left[\log(\lambda_j^2) - \frac{\beta_j^2}{\lambda_j^2} + 2\frac{\lambda_j^2}{\tau^2}\left(1 + \frac{\lambda_j^2}{\tau^2}\right)^{-1}\right]$$

$$\frac{\partial^2}{(\partial \log(\lambda_j^2))^2} \log(p(\log(\lambda_j^2)|\boldsymbol{\beta}, \tau, \boldsymbol{y}, \boldsymbol{x})) = -\frac{\partial}{\partial \log(\lambda_j^2)}\frac{1}{2}\sum_{i=1}^{n}\frac{\frac{\partial}{\partial \log(\lambda_j^2)} s_i^2}{s_i^2} - \frac{1}{2}\sum_{i=1}^{n}\frac{1}{2}\boldsymbol{z}'\frac{\partial^2}{(\partial \log(\lambda_j^2))^2}(SS')^{-1}\boldsymbol{z}$$

$$+\boldsymbol{\beta}'B'\sum_{i=1}^{n}\frac{\partial^2}{(\partial \log(\lambda_j^2))^2}S^{-1}\boldsymbol{z} - \frac{1}{2}\left[\log(\lambda_j^2) - \frac{\beta_j^2}{\lambda_j^2} + 2\frac{\lambda_j^2}{\tau^2}\left(1 + \frac{\lambda_j^2}{\tau^2}\right)^{-1}\right] \tag{14}$$

and for $\log(\tau)$

$$\frac{\partial}{\partial \log(\tau)}\log(p(\log(\tau)|\boldsymbol{\lambda}, \boldsymbol{y}, \boldsymbol{x})) = -(p-1) - \frac{2\tau^2}{1+\tau^2} + \sum_{j=1}^{p}\frac{\lambda_j^2}{\tau^2}(1+\frac{\lambda_j^2}{\tau^2})^{-1}$$

$$\frac{\partial^2}{(\partial \log(\tau))^2}\log(p(\log(\tau)|\boldsymbol{\lambda}, \boldsymbol{y}, \boldsymbol{x})) = \frac{4\tau^4}{(1+\tau^2)^2} - \frac{4\tau^2}{(1+\tau^2)} + \sum_{j=1}^{p}(\frac{2\lambda_j^2}{\tau^2})^2(1+\frac{\lambda_j^2}{\tau^2})^{-2}$$

$$-\sum_{j=1}^{p}\frac{4\lambda_j^2}{\tau^2}(1+\frac{\lambda_j^2}{\tau^2})^{-1}.$$

To arrive at (14), note that the first two derivatives of $s_i^2$ with respect to $\log(\lambda_j^2)$ are

$$\frac{\partial}{\partial \log(\lambda_j^2)}s_i^2 = -b_{ij}^2\lambda_j^2(1 + \sum_{j=1}^{p}b_{ij}^2\lambda_j^2)^{-2}$$

$$\frac{\partial^2}{(\partial \log(\lambda_j^2))^2}s_i^2 = -b_{ij}^2\lambda_j^2(1 + \sum_{j=1}^{p}b_{ij}^2\lambda_j^2)^{-2} + (b_{ij}^2\lambda_j^2)^2(1 + \sum_{j=1}^{p}b_{ij}^2\lambda_j^2)^{-3},$$

where

$$s_i^2 = (1 + \boldsymbol{b}_i'\operatorname{diag}(\boldsymbol{\lambda})^2\boldsymbol{b}_i)^{-1} = (1 + \sum_{j=1}^{p}b_{ij}^2\lambda_j^2)^{-1}.$$

## B.3 Bayesian Variable Selection Implicit Copula

Generation of the indicators $(\gamma_i, \gamma_j)$ for estimating the BVSC requires evaluation of $A(\gamma_i, \gamma_j)$ for all four configurations $(\gamma_i, \gamma_j) \in \{(0,0), (0,1), (1,0), (1,1)\}$, which can be undertaken efficiently as follows.

For $k \in \{0, 1\}$, let $\boldsymbol{\gamma}^{(k)} = (\gamma_1, \ldots, \gamma_{l-1}, k, \gamma_{l+1}, \ldots, \gamma_p)$, for an element $1 \leq l \leq p$, $U_k$ be an upper triangular Cholesky factor such that $U_k'U_k = B_{\boldsymbol{\gamma}^{(k)}}'B_{\boldsymbol{\gamma}^{(k)}}$, and $M_k = B_{\boldsymbol{\gamma}^{(k)}}U_k^{-1}$ be an $n \times p_{\gamma^{(k)}}$ matrix. If $\gamma_l = 0$, then $U_1$ can be readily computed from $U_0$ using Cholesky updating in $O(p_{\gamma^{(1)}}^2)$ operations, and $M_1$ evaluated by solving the system of $M_1 U_1 = B_{\gamma^{(1)}}$. Similarly, if $\gamma_l = 1$, then $U_0$ can be readily computed from $U_1$ using Cholesky down-dating in $O(p_{\gamma^{(0)}}^2)$ operations, and $M_0$ evaluated by solving the system of $M_0 U_0 = B_{\gamma^{(0)}}$.

These relationships allow rapid computation of the Cholesky factor $U$ (such that $U'U = B_\gamma' B_\gamma$) and $M = B_\gamma U^{-1}$ for each of the four configurations of $\gamma$, because they only differ by up to two elements. Given these matrices, for each configuration we compute:

(i) $s_i = (1 + c \sum_{j=1}^{p_\gamma} m_{ij}^2)^{-1/2}$ for $i = 1, \ldots, n$, and set $\tilde{z} = (\tilde{z}_1, \ldots, \tilde{z}_n)'$, where $\tilde{z}_i = z_i/s_i$ and $M = \{m_{ij}\}$;

(ii) $z' R(x, \gamma)^{-1} z = \tilde{z}'(I - \frac{c}{1+c} MM')\tilde{z}$ by solving $\zeta = M'\tilde{z}$; and,

(iii) $|R(x, \gamma)| = (\prod_{i=1}^n s_i^2)|I + cMM'| = (\prod_{i=1}^n s_i^2)|I + cM'M| = (\prod_{i=1}^n s_i^2)(1+c)^{p_\gamma}$, where we have used Sylvester's determinant identity.

At no stage is either $R$ or $MM' = B_\gamma (B_\gamma' B_\gamma)^{-1} B_\gamma'$ computed directly, which would be prohibitive because they are both $n \times n$ matrices. From the terms at (i)–(iii) above, it is straightforward to evaluate $A(\gamma_i, \gamma_j)$ for all four configurations.

## Appendix C  Additive PSC Sampling Scheme

Below is one sweep of a MCMC scheme for simulating from the posterior of the copula smoother with an additive PSC, given the values $z_i = \Phi^{-1}(\hat{F}_{Y_i}(y_i))$ for $i = 1, \ldots, n$.

**Sampling Scheme** *(One Sweep)*

For $l = 1, \ldots, L$:

  Step 1. Generate from $p(\beta_l | \{\beta_k; k \neq l\}, \theta, y) = p(\beta_l | \{\beta_k; k \neq l\}, \theta, z)$

  Step 2. Generate from $p(\theta_l | \beta, \{\theta_k; k \neq l\}, y) = p(\theta_l | \beta, \{\theta_k; k \neq l\}, z)$

  Step 3. Update $S = \text{diag}(s_1, \ldots, s_n)$

At Step 1, when $l = 1, \ldots, L-1$, $\beta_l \sim N(\mu_{\beta_l}, \Sigma_{\beta_l})$ constrained so that $I(\mathbf{1}'B_l\beta_l = 0)$, where $\Sigma_{\beta_l} = (B_l'B_l + P_l(\theta_l))^{-1}$, $\mu_{\beta_l} = \Sigma_{\beta_l} B_l' S^{-1} e_l$ and $e_l = z - \sum_{k \neq l} B_k \beta_k$. To implement this step at sweep $j$, $e_l$ can be computed using the fast updating formula $e_l = e_{l-1} - B_l \beta_l^{(j-1)} + B_{l-1} \beta_{l-1}^{(j)}$, where $\beta_l^{(j)}$ is the value of $\beta_l$ generated at sweep $j$. To generate from a linearly constrained Gaussian we use Algorithm 2.6 of Rue and Held (2005). When $l = L$, the distribution is unconstrained.

At Step 2, $\boldsymbol{\theta}_l$ is generated using a Metropolis-Hastings step in a similar fashion as for the univariate model. Step 3 updates each element $s_i = (1 + \sum_{l=1}^{L} \mathcal{Q}_{il})^{-1/2}$, where $\mathcal{Q}_{il} = \boldsymbol{b}'_{il} P(\boldsymbol{\theta})^{-1} \boldsymbol{b}_{il}$, for $i = 1, \ldots, n$. This can be computed efficiently by storing the values $\{\mathcal{Q}_{il}\}_{i=1:n;l=1:L}$, and only updating $\mathcal{Q}_{1l}, \ldots, \mathcal{Q}_{nl}$ at Step 3. As is usual with P-splines, implementation involves Cholesky factorization of $(B'_l B_l + P_l(\boldsymbol{\theta}_l))$ and $P_l(\boldsymbol{\theta}_l)$, both of which are banded matrices and fast to factor.

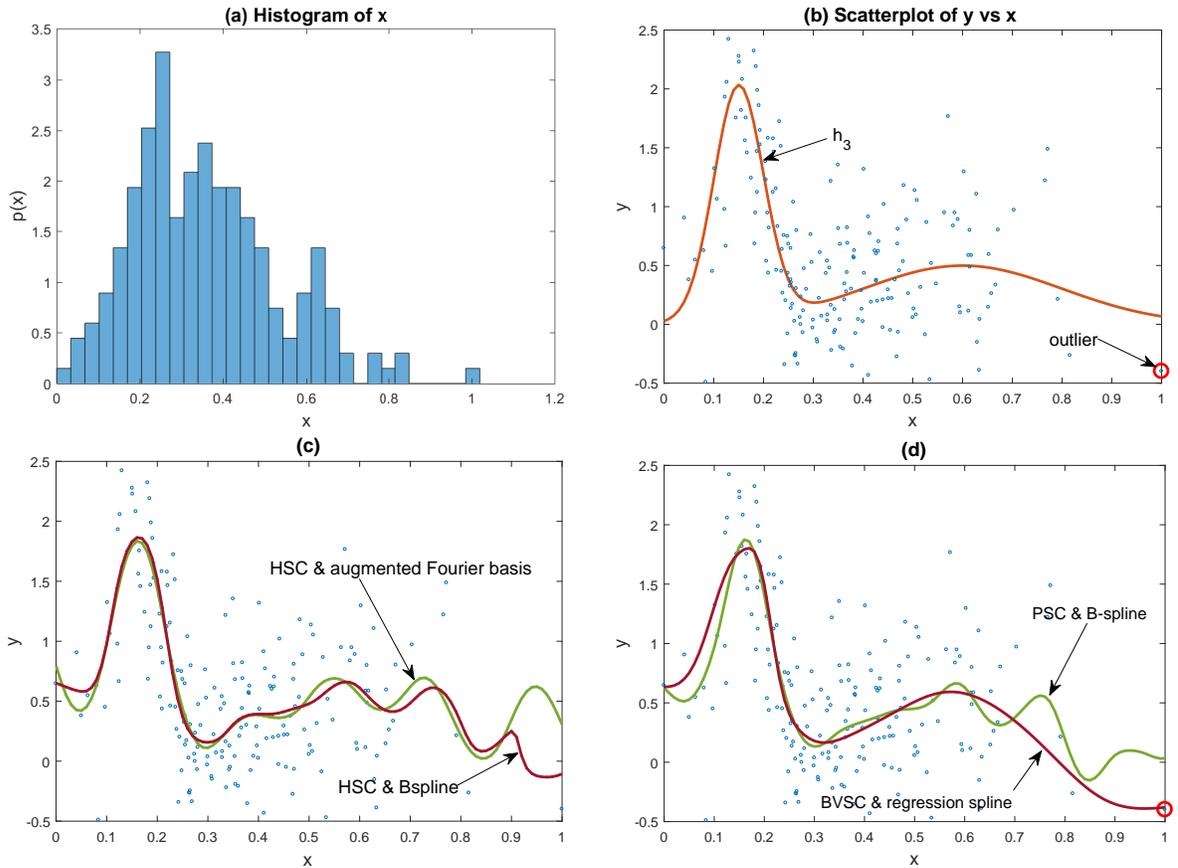

Figure 1: Summary of the illustrative dataset with $n = 200$ observations. Panel (a) plots a histogram of the covariate values $x_1, \ldots, x_n$ which were generated from a chi-square distribution scaled to $[0, 1]$. Panel (b) contains a scatterplot of $x_i$ versus $y_i \sim N(h_3(x_i), 0.5^2)$, with the function $h_3$ plotted as a red line. Panels (c,d) contain the posterior mean function estimates $\hat{f}$ from the four copula models fit to the data.

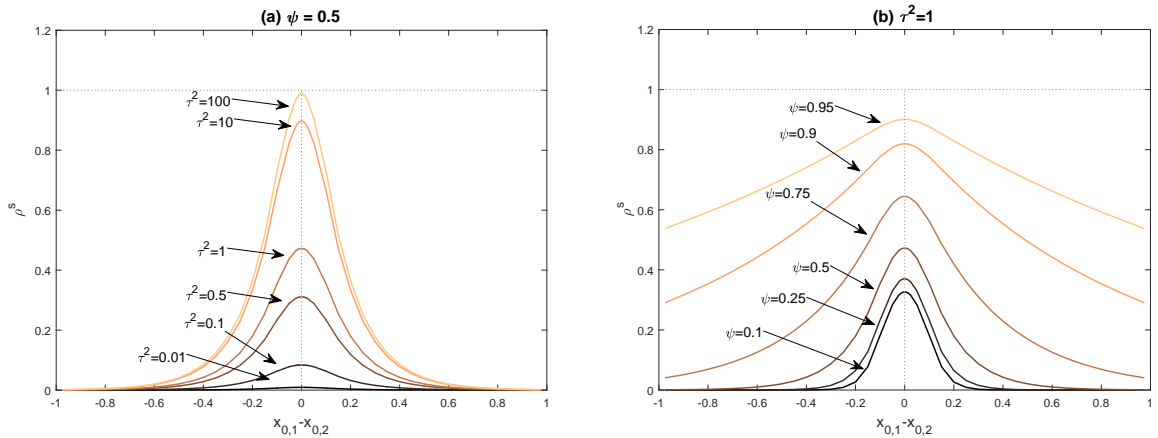

Figure 2: Spearman's rho $\rho^s(x_{0,1}, x_{0,2}|\boldsymbol{x}, \boldsymbol{\theta})$ plotted against $(x_{0,1} - x_{0,2})$ for the PSC with B-spline basis and conditional on $\boldsymbol{\theta}$. In panel (a), $\psi = 0.5$ and $\tau^2 \in \{0.01, 0.1, 0.5, 1, 10, 100\}$. In panel (b), $\tau^2 = 1$ and $\psi \in \{0.1, 0.25, 0.5, 0.75, 0.9, 0.95\}$.

| Implicit Copula | Parameters | $P(\boldsymbol{\theta})$ | $s_i$ | $R(\boldsymbol{x},\boldsymbol{\theta},\boldsymbol{\gamma})$ | Suggested Bases |
|---|---|---|---|---|---|
| PSC | $\boldsymbol{\theta}=\{\tau^2,\psi\}$ <br> $\boldsymbol{\gamma}=\emptyset$ | $\frac{1}{\tau^2}P(\psi)$ | $(1+\tau^2 \boldsymbol{b}_i' P(\psi)^{-1}\boldsymbol{b}_i)^{-1/2}$ | $S(\boldsymbol{x},\boldsymbol{\theta})(I+\tau^2 B P(\psi)^{-1}B')S(\boldsymbol{x},\boldsymbol{\theta})'$ | B-Splines |
| HSC | $\boldsymbol{\theta}=\{\boldsymbol{\lambda},\tau\}$ <br> $\boldsymbol{\gamma}=\emptyset$ | $\mathrm{diag}(\boldsymbol{\lambda})^{-2}$ | $(1+\boldsymbol{b}_i' \mathrm{diag}(\boldsymbol{\lambda})^2 \boldsymbol{b}_i)^{-1/2}$ | $S(\boldsymbol{x},\boldsymbol{\theta})(I+B\,\mathrm{diag}(\boldsymbol{\lambda})^2 B')S(\boldsymbol{x},\boldsymbol{\theta})'$ | Augmented Fourier/ Radial Bases |
| BVSC | $\boldsymbol{\theta}=\emptyset$ <br> $\boldsymbol{\gamma}$ | $\frac{1}{c}B_{\boldsymbol{\gamma}}'B_{\boldsymbol{\gamma}}$ | $(1+c\boldsymbol{b}_{\boldsymbol{\gamma},i}'(B_{\boldsymbol{\gamma}}'B_{\boldsymbol{\gamma}})^{-1}\boldsymbol{b}_{\boldsymbol{\gamma},i})^{-1/2}$ | $S(\boldsymbol{x},\boldsymbol{\gamma})(I+cB_{\boldsymbol{\gamma}}(B_{\boldsymbol{\gamma}}'B_{\boldsymbol{\gamma}})^{-1}B_{\boldsymbol{\gamma}}')S(\boldsymbol{x},\boldsymbol{\gamma})'$ | Regression Splines, Radial Bases |

Table 1: Summaries of the implicit copulas constructed from the three regularized regression smoothers, along with suggested matching bases. The elements of the matrix $S(\boldsymbol{x},\boldsymbol{\theta},\boldsymbol{\gamma}) = \mathrm{diag}(s_1,\ldots,s_n)$ are defined in the fourth column.

| Functional Relationship | Posterior Predictive Mean | MCMC Estimator |
|---|---|---|
| $\tilde{m}(x_0) = \mathbb{E}(\tilde{Z}_0\|x_0,\boldsymbol{\beta}) = \boldsymbol{b}_0'\boldsymbol{\beta}$ | N/A | N/A |
| $m(x_0) = \mathbb{E}(Z_0\|x_0,\boldsymbol{x},\boldsymbol{\beta},\boldsymbol{\theta},\boldsymbol{\gamma}) = s_0\boldsymbol{b}_0'\boldsymbol{\beta}$, where $s_0 = (1 + \boldsymbol{b}_0'P(\boldsymbol{\theta})^{-1}\boldsymbol{b}_0)^{-1/2}$ | $\mathbb{E}(Z_0\|x_0,\boldsymbol{x},\boldsymbol{y})$ $= \int m(x_0)p(\boldsymbol{\beta},\boldsymbol{\theta},\boldsymbol{\gamma}\|\boldsymbol{x},\boldsymbol{y})\mathrm{d}(\boldsymbol{\beta},\boldsymbol{\theta},\boldsymbol{\gamma})$ | $\hat{m}(x_0) = \boldsymbol{b}_0'\left(\frac{1}{J}\sum_{j=1}^{J} s_0^{[j]}\boldsymbol{\beta}^{[j]}\right)$, where $s_0^{[j]} = (1 + \boldsymbol{b}_0'P(\boldsymbol{\theta}^{[j]})^{-1}\boldsymbol{b}_0)^{-1/2}$ |
| $f(x_0) = \mathbb{E}(Y_0\|x_0)$ | $\mathbb{E}(Y_0\|x_0,\boldsymbol{x},\boldsymbol{y}) = \int \mathbb{E}(Y_0\|x_0,\boldsymbol{x},\boldsymbol{\beta},\boldsymbol{\theta},\boldsymbol{\gamma})$ $\times p(\boldsymbol{\beta},\boldsymbol{\theta},\boldsymbol{\gamma}\|\boldsymbol{x},\boldsymbol{y})\mathrm{d}(\boldsymbol{\beta},\boldsymbol{\theta},\boldsymbol{\gamma})$ | 1. $\hat{f}(x_0) = \frac{1}{J}\sum\left\{\int \hat{F}_Y^{-1}(\Phi_1(z_0))\phi_1\left(\frac{z_0 - m_0^{[j]}(x_0)}{s_0^{[j]}}\right)\mathrm{d}z_0\right\}$, where $m_0^{[j]}(x_0) = s_0^{[j]}(x_0)\boldsymbol{b}_0'\boldsymbol{\beta}^{[j]}$ 2. $\tilde{f}(x_0) = \int \hat{F}_Y^{-1}(\Phi_1(z_0))\phi_1\left(\frac{z_0 - \hat{m}_0(x_0)}{\hat{s}_0}\right)\mathrm{d}z_0$, where $\hat{s}_0 = \frac{1}{J}\sum_{j=1}^{J} s_0^{[j]}$ |

Table 2: Summary of the functional relationships in the implicit copula model, along with their Bayesian posterior predictive means and estimators. The estimators employ the marginal distribution estimate $\hat{F}_Y$, and the output of the MCMC scheme $\{\boldsymbol{\beta}^{[j]},\boldsymbol{\theta}^{[j]},\boldsymbol{\gamma}^{[j]}; j = 1,\ldots,J\}$. The final row gives the estimators for the regression function from the copula smoother.

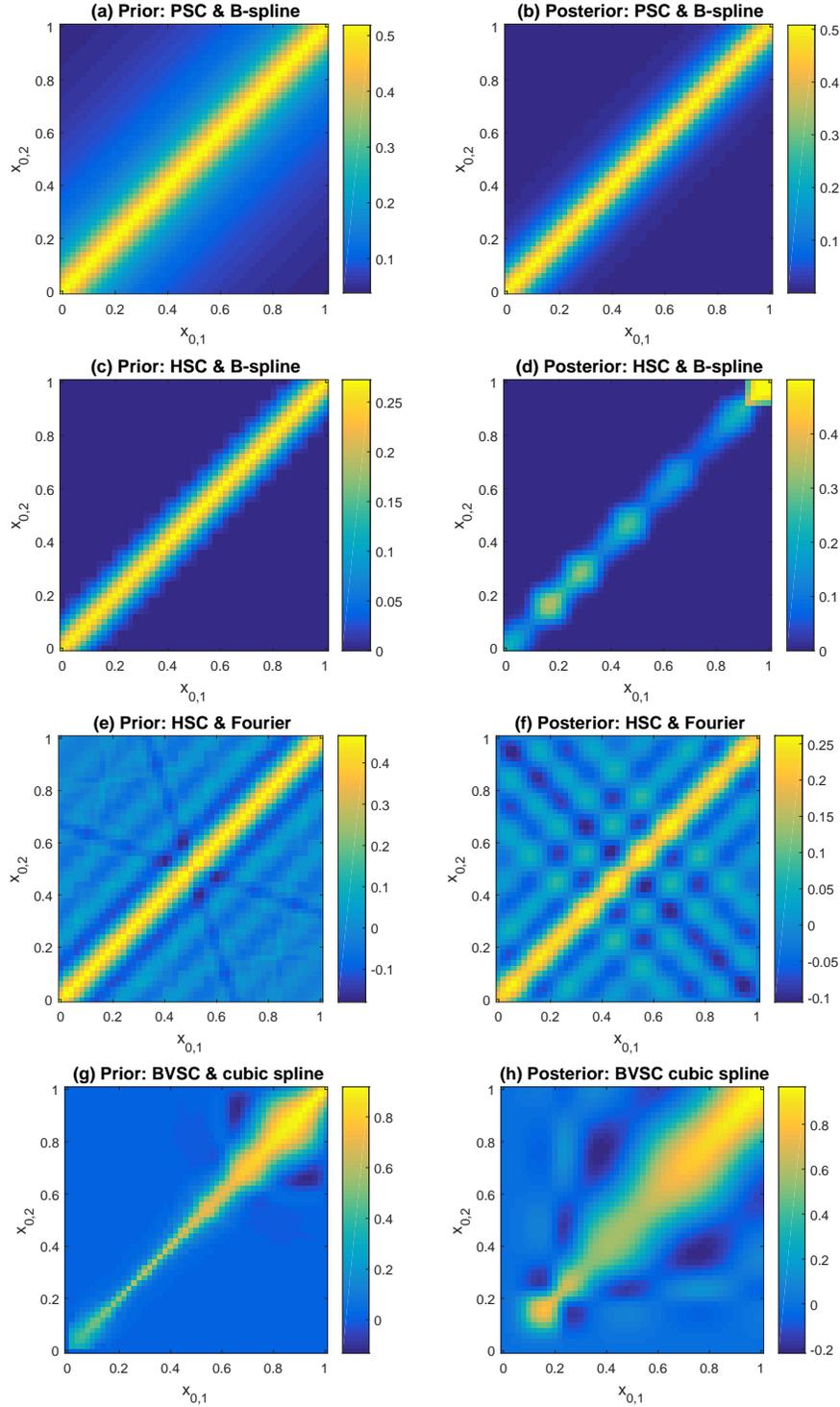

Figure 3: Bivariate surfaces of Spearman's rho $\rho^s_\pi(x_{0,1}, x_{0,2}|\boldsymbol{x})$ of $C_\pi$ over $(x_{0,1}, x_{0,2}) \in [0,1]^2$. The left column gives results when $(\boldsymbol{\theta}, \boldsymbol{\gamma})$ is integrated with respect to the prior $\pi_0$. The right column gives results when $(\boldsymbol{\theta}, \boldsymbol{\gamma})$ is integrated with respect to the posterior using the data in Figure 1. The panels give results for different shrinkage prior/basis combinations: (a,b) PSC/B-spline; (c,d) HSC/B-spline; (e,f) HSC/augmented Fourier basis; (g,h) BVSC/regression spline.

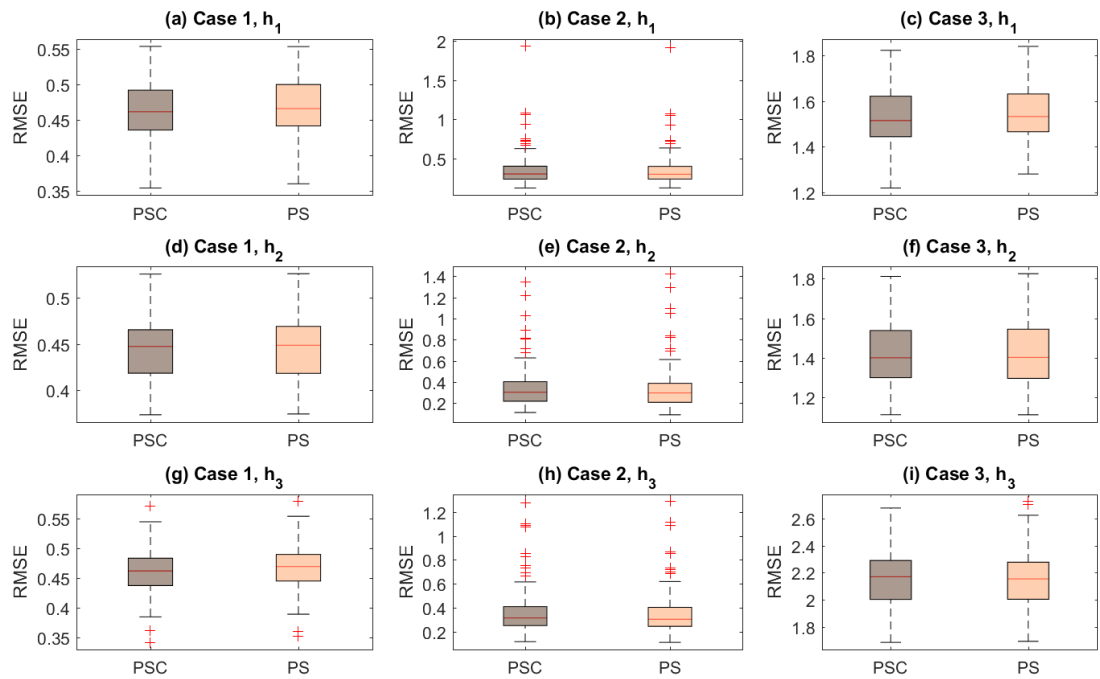

Figure 4: Comparison of root mean square error values from the simulation study. Each panel corresponds to a different combination of test function and case. The first column (a,d,g) is for Case 1, the second column (b,e,h) is for Case 2, and the third column (c,f,i) is for Case 3. The first row (a,b,c) is for $h_1$, the second row (d,e,f) is for $h_2$ and the third row (g,h,i) is for $h_3$. Each boxplot is of the 100 values of the RMSE$(f_j, \hat{f}_j)$ metric from the simulation replicates. The left-hand boxplot is for the PSC estimator, and the right-hand boxplot for the PS estimator. Lower values correspond to increased accuracy.

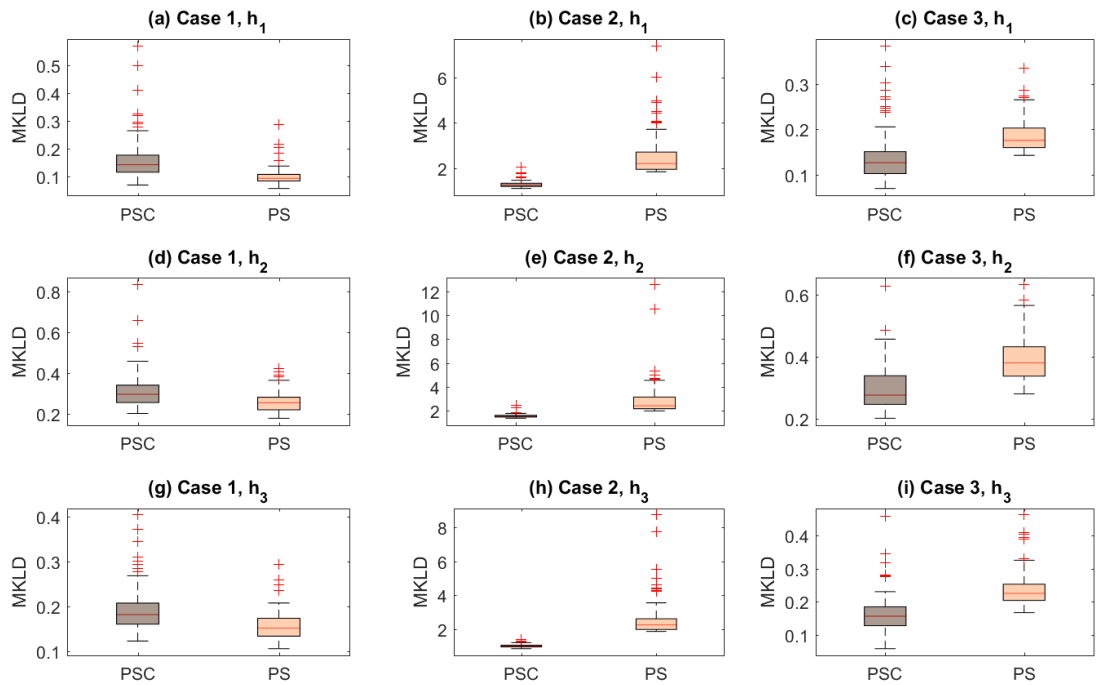

Figure 5: Comparison of mean Kullback-Leibler divergence values from the simulation study. The panels are arranged as outlined in the caption to Figure 4. Each boxplot is of the 100 values of the MKLD$(p||\hat{p})$ metric from the simulation replicates. The left-hand boxplot is for the PSC estimator, and the right-hand boxplot for the PS estimator. Lower values correspond to increased accuracy.

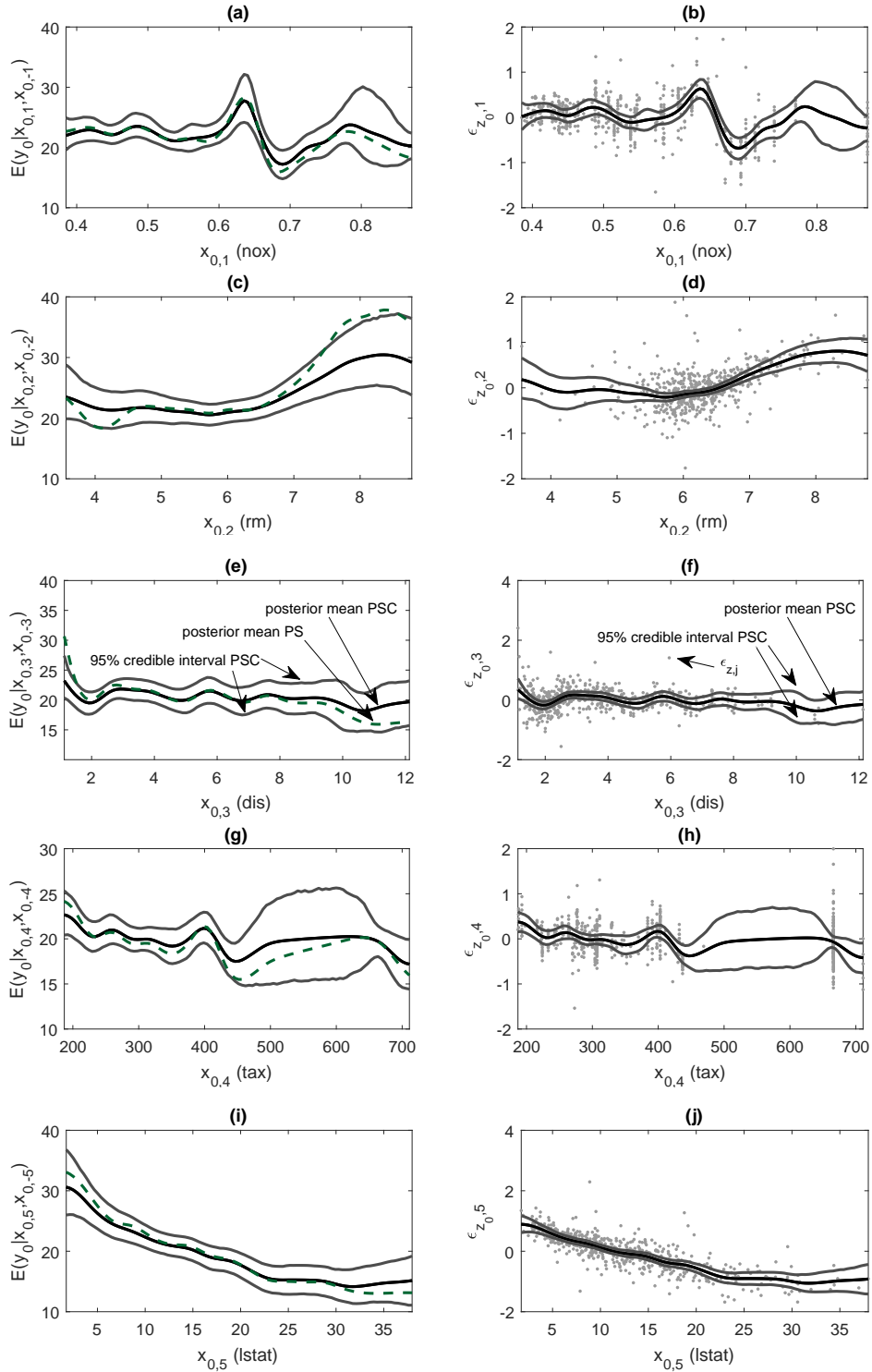

Figure 6: Summary of the copula smoother with an additive PSC fitted to the Boston housing data. The left panels plot slices of the estimated regression surface $\hat{f}$ for each of (a) NOX, (c) RM, (e) DIS, (g) TAX, and (e) LSTAT, fixing the other four covariate values to those of the median priced house. Estimates are given for both the copula smoother (bold line) and additive PS (dashed line) for comparison. The 95% credible intervals are also given for the copula smoother. The right panels plot $\hat{m}_l$ and the partial residuals $\{\epsilon_{1,l}, \ldots, \epsilon_{n,l}\}$ for (b) NOX $l=1$, (d) RM $l=2$, (f) DIS $l=3$, (h) TAX $l=4$, and (j) LSTAT $l=5$.

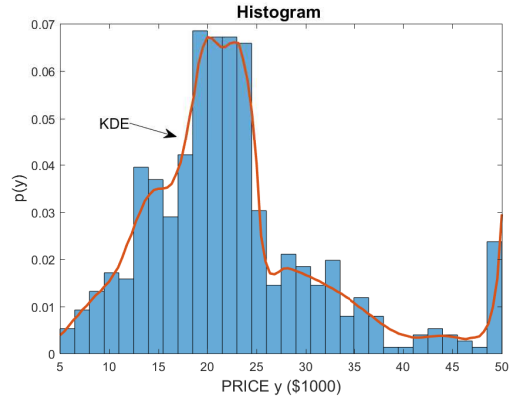

Figure 7: Histogram of PRICE (in $1,000) in the Boston housing dataset. Also shown in red is the adaptive kernel density estimate (KDE).

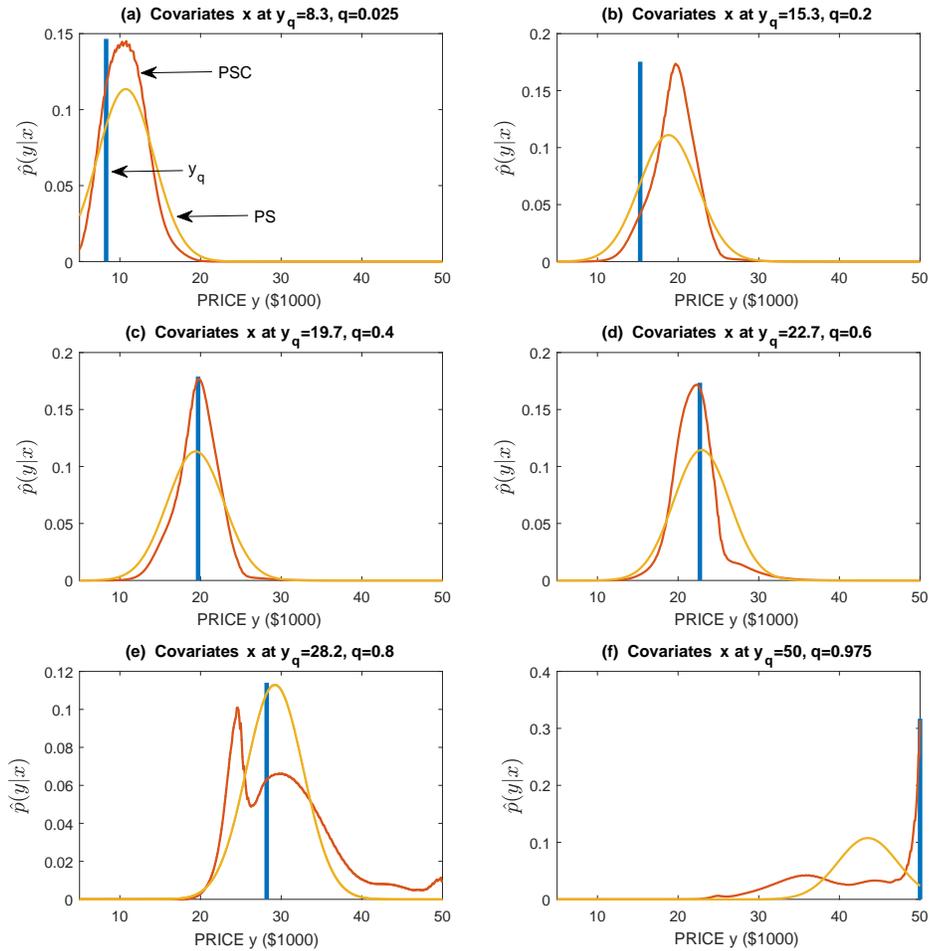

Figure 8: Predictive densities $\hat{p}(y|\boldsymbol{x})$ for six houses in the Boston housing data. Each corresponds to the house at the $q$th quantile of the observed prices, for (a) $q = 0.025$, (b) $q = 0.2$, (c) $q = 0.4$, (d) $q = 0.6$, (e) $q = 0.8$ and (f) $q = 0.975$. In each panel the predictive density is plotted for the additive copula smoother (red line) and the Gaussian P-spline (yellow line), while the observed price is marked with a blue vertical line.

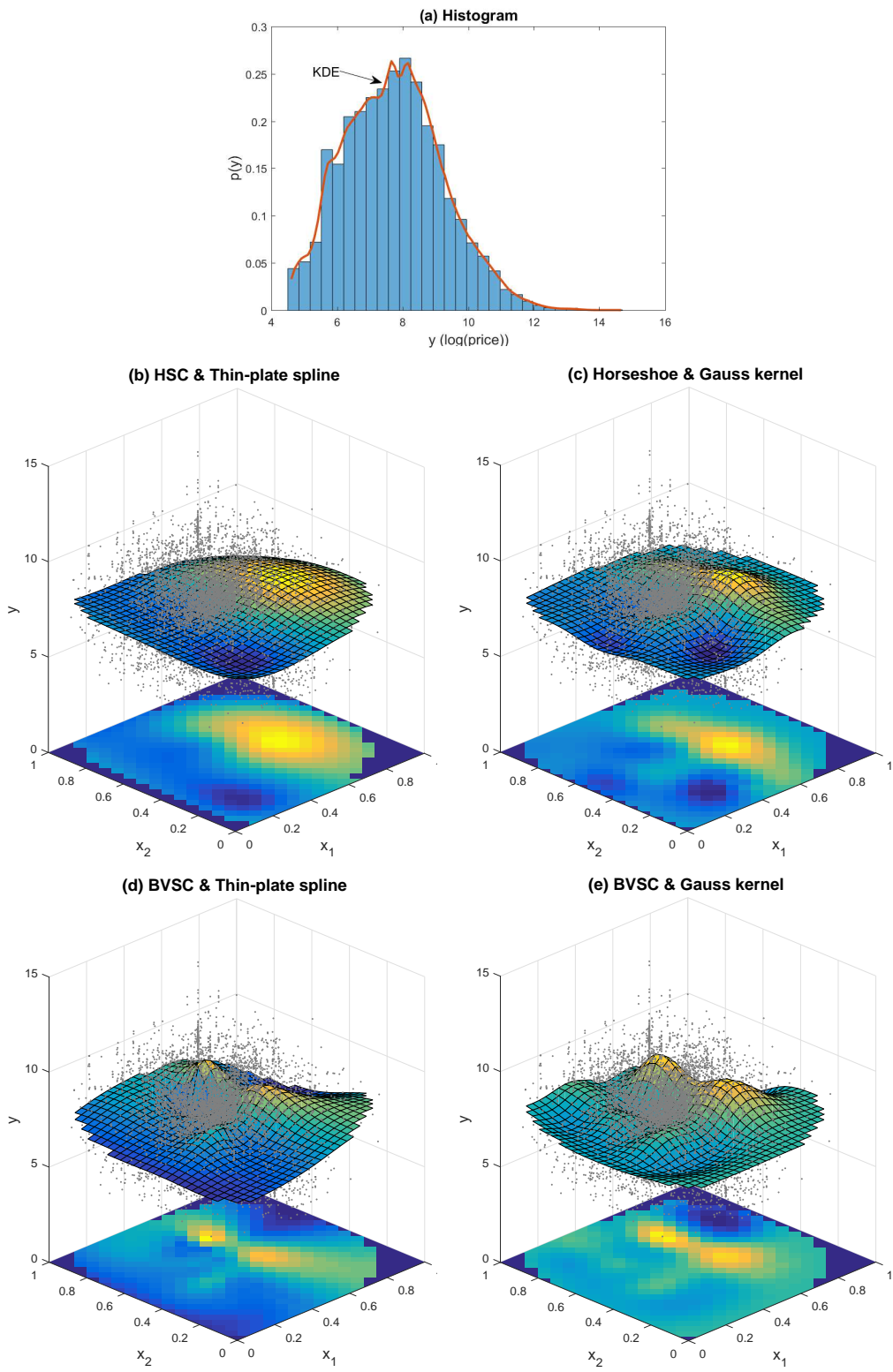

Figure 9: Bivariate surface estimates for the print price data. Panel (a) gives a histogram and KDE of the logarithm of the sale price $Y$. The remaining four panels show the bivariate function estimates $\hat{f}(\boldsymbol{x})$ for the copula smoother with four shrinkage prior/radial basis combinations: (b) HSC & thin-plate spline, (c) HSC & Gaussian kernel, (d) BVSC & thin-plate spline, (e) BVSC & Gaussian kernel.